\documentclass{aa}  

\usepackage{graphicx}
\usepackage{txfonts}
\usepackage{lipsum}
\usepackage{subcaption}
\usepackage{lscape}
\usepackage{placeins}
\usepackage{hyperref}

\usepackage{xcolor}
\usepackage{lineno}
\usepackage{tikz}
\usepackage{dsfont}
\usepackage{enumitem}

\begin{document}

\newcommand{\psj}{PSJ}

\titlerunning{Catastrophic tidal disruption of heterogeneous rubble piles}
\title{Catastrophic tidal disruption of heterogeneous rubble piles: a tale of two regimes}

\author{John Wimarsson\inst{1}\corrauth{john.wimarsson@unibe.ch}
\and Eric Frizzell\inst{2}\email{ericscott.frizzell@polimi.it}
\and Martin Jutzi\inst{1}\email{martin.jutzi@unibe.ch}
\and Fabio Ferrari\inst{2,1}\email{fabio1.ferrari@polimi.it}
}

\institute{Space Research \& Planetary Sciences, Physics Institute, University of Bern,
Gesellschaftsstrasse 6, 3012, Bern, Switzerland
\and Department of Aerospace Science and Technology, Politecnico di Milano,
20156, Milano, Italy}

 \abstract
   {The way a rubble-pile body deforms or disrupts under the influence of tidal forces can be directly tied to its internal strength and configuration. Computational modelling of such tidal disruption events provides an indispensable numerical laboratory for constraining the origin and evolution of small bodies in the Solar System.}
   {A majority of previous investigations into tidal disruption of rubble piles have mainly considered progenitors consisting of same-sized, spherical elements. Our study attempts to fill the existing gap in studies analysing the effect of aggregate heterogeneity on tidal disruption outcomes by varying element shape and size frequency distribution. Such heterogeneities have been shown to strongly influence rubble pile dynamics for impacts and rotational failure.} 
   {We performed over one hundred numerical simulations of parabolic and hyperbolic tidal encounters between six unique rubble-pile progenitors and the Earth using the \textit{N}-body code GRAINS. The resulting mass distributions of generated fragments and tidal chain morphologies for the different progenitors were further tied to the internal strength of rubble piles.}
   {Two regimes of tidal disruption are identified. In the first regime, closest to the planet, the dynamic evolution is dominated by tidal forces. Here, particle shape, size distribution and resolution appear to have little importance for the resulting distribution of fragment masses. In the second, shear-controlled regime, the internal structure of the progenitor begins to strongly influence the resulting tidal chain morphology and properties of the surviving fragments.}
   {Heterogeneity originating from the shape and size frequency distribution of elements in rubble pile models has a substantial effect on the outcomes of tidal disruption events. These parameters must be carefully taken into account when future studies attempt to tie results from numerical models to observations.}


\maketitle
\nolinenumbers

\section{Introduction}\label{section:introduction}

Any small body undergoing close encounters with a larger object, such as a planet or star, will be subjected to significant tidal forces. For structurally weak objects like rubble piles \citep{walsh2018}, these events can lead to surface modifications \citep{binzel2010,yu2014}, distortion \citep{richardson1998,bottke1999} or even complete disruption, as in the case of comet Shoemaker-Levy 9 that was torn into 21 detected fragments when it passed Jupiter within 1.33 Jovian radii in 1992 \citep{sekanina1994,asphaug1994}. The resulting debris from such events can in turn form stable binary asteroid systems \citep{walsh2006}, or in the case of white dwarfs, create observable debris discs \citep{jura2003,li2021}. Catastrophic encounters with terrestrial planets provide a solid explanation for discrepancies in the predicted and observed populations of near-Earth objects (NEOs) \citep{granvik2016,granvik2024}. They have further been postulated as a source of NEO families \citep{schunova2014}. Hence, understanding the fundamental underlying dynamics of tidal disruptions and how the outcome relates to the internal structure of a rubble pile is key to deepening our knowledge of the origin and evolution of small bodies in planetary systems. 

Given the nearly strengthless nature of rubble pile objects \citep{richardson2002}, even slight variations in their internal structure can considerably alter the way they deform or disrupt. Introducing heterogeneities such as larger boulders or a solid core in numerical models can, for example, prevent a rubble pile from collapsing under high spin rates \citep{hirabayashi2015_core,zhang2017,zhang2018,zhang2021,sanchez2018,ferrari&tanga2022} or due to impacts \citep{raducan&jutzi2022,raducan2024b}. Similar effects are seen for gravitational aggregates with low levels of interparticle cohesion \citep{hirabayashi2015_cohesion,sanchez2014,zhang2018,raducan2024a}. Another form of heterogeneity that notably increases the internal strength of rubble piles is irregularities in particle shape. To accurately capture the dynamics of granular media, it is crucial to resolve mechanical effects such as interlocking, polyhedral contacts, dilatancy\footnote{The resulting increase in volume of a granular medium when exposed to shear deformation.}, particle spin orientation and off-centre collisions. As a result, utilising non-spherical elements in \textit{N}-body simulations of rubble piles substantially increases structural stability \citep{korycansky2006,korycansky2009,ferrari&tanga2020,marohnic2023}.

In this work, we use discrete element method (DEM) \textit{N}-body simulations to highlight the influence of heterogeneities in models of rubble-pile objects during catastrophic tidal disruption events, showing that particle shape and size frequency distribution strongly affect the resulting length and morphology of tidal chains. Moreover, we demonstrate that observed tidal chains, such as the one produced by Shoemaker-Levy 9, are best characterised with mass frequency distributions rather than fragment number due to the stochastic nature of the clustering process. In Sect.~\ref{section:previous_work}, we briefly summarise previous work on tidal encounters of rubble piles. We then present the numerical methods used in our simulations, as well as the analysis in Sect.~\ref{section:method}. Section~\ref{section:results_fiducial} shows the outcomes of the fiducial set of simulations, focusing on the distribution of masses for the resulting remnants. These results are compared to the corresponding mass distributions and tidal chain morphologies for alternative homogeneous and heterogeneous aggregates in Sect.~\ref{section:results_heterogeneity}. Finally, we provide conclusions and an outlook for future studies in Sect.~\ref{section:conclusions}. For the remainder of this work, we refer to a collection of particles as an aggregate or rubble pile, while the initial configuration of boulders is called a progenitor. For the post-disruption aggregates, we use fragment and remnant interchangeably.

\section{Previous work} \label{section:previous_work}

To a first approximation, the disruption of a self-gravitating body during a close encounter can be estimated using continuum theory. The distance at which a fully fluid body structurally fails is known as the fluid Roche limit, $d_\mathrm{FRL}$, after the analysis by \cite{roche1847}. For a spherical object of density $\rho_m$ approaching a significantly larger body of density $\rho_M$ and radius $R$, the distance is approximately given by $d_\mathrm{FRL} = 2.44R (\rho_M/\rho_m)^{1/3}$. As rubble piles have non-negligible internal strength, the fluid treatment leads to greatly overestimated distances, which has caused many studies to revisit the theory \citep[e.g.][]{aggarwal1974,sridhar1992}. For reference, the rigid Roche limit is $d_\mathrm{RRL}=(2\rho_M/\rho_m)^{1/3}$. Using static theory, \citet{holsapple2006,holsapple2008} made an attempt to arrive at a better indicator for rubble pile disruption, accounting for the angle of friction and cohesion of the granular material, as well as physical aspect ratios. Even so, analytical solutions built on continuum theory remain unable to resolve the characteristic particle--particle interactions of granular media, which introduce complex dynamical effects.

The discrepancy between predicted disruption distances and the corresponding values identified in simulations, and the related influence of internal structure, have been studied and discussed in several works. Initially, \citet{asphaug1994,asphaug1996} employed DEM \textit{N}-body simulations with frictionless progenitors consisting of equal-sized spheres to estimate the size, density and internal structure of Shoemaker-Levy 9 (SL9), showing that their progenitors were disrupted far inside the fluid Roche limit. Later studies such as \citet{richardson1998} opted for codes including friction, using non-deformable hard-sphere DEM contacts with hexagonal close packing. Their results emphasised that higher elongations and spin rates of the progenitor makes it more susceptible to disruption. Here, the authors also defined three regimes of disruption: S-class, where the largest remnant has less than 50\% of the initial mass; B-class, where the largest remnant has between 50\% and 90\% of the mass; M-class, where the progenitor retains more than 90\% of its mass. The topic is revisited by \citet{zhang2020b}, using a more modern contact model with deformable, equal-sized soft spheres and parametrised non-spherical contacts, showing that such an implementation leads to rubble pile models that are more resistant to structural failure. The authors further conclude that random packing results in weaker structures compared to hexagonal close packing. From the disruption outcomes, it is also clear that their soft-sphere model produces disruption patterns more comparable to the hard-contact polyhedral approach that had earlier been used by \citet{movshovitz2012}, arriving at similar estimates of the density of SL9. The model by \citet{movshovitz2012} was the first tidal disruption study to introduce heterogeneities by employing irregularly shaped particles with a uniform size frequency distribution (SFD). Furthermore, \citet{marohnic2026} recently expands on the use of non-spherical particles by employing a method where each aggregate constituent is made up of glued-together spheres with the same contact model as \citet{zhang2020b}, finding a significant increase in rubble pile stability compared to models consisting of single soft spheres. Extending the scope beyond hyperbolic encounters with terrestrial planets, the single soft-sphere methodology is further combined with a polydisperse SFD to simulate close encounters between a rubble pile and a main sequence star in \citet{zhang2020a}. The same model is also compared with analytic estimates to determine the tidal failure modes of the Martian moon Phobos by \citet{agrusa2026}, who find that rubble-pile satellites can undergo significant tidal stripping before they are fully disrupted.

It is difficult to directly compare the results of the aforementioned DEM studies given the spread in: numerical implementations; the variation in particle sizes, shapes and packing; contact methods and material properties; as well as the numerical capabilities available at the time each study was conducted. Nevertheless, the general trends and conclusions appear to hold. Heterogeneities in the form of irregular particle shapes combined with soft contact methods substantially increase the structural integrity of rubble piles during tidal encounters. With the notable effects of granular mechanisms in mind, we also acknowledge that polydispersity remains a largely unexplored, additional degree of heterogeneity for tidal disruption simulations, while it has, as mentioned, been shown to greatly influence the stability of rubble piles in the scope of rotational failure and impacts. Additionally, observations of boulders on asteroid surfaces indicate that rubble-pile constituent SFDs are best fit with power laws and exponential functions \citep{pajola2024}. In turn, the core goal of this study is to continue building upon the solid foundation of results from previous works and introduce this additional degree of heterogeneity, comparing the effect for both spheres and irregularly shaped, polyhedral elements using an advanced contact model.

\section{Numerical method}\label{section:method}

\subsection{\textit{N}-body modelling}

To perform the simulations in this work, we made use of the discrete element method \textit{N}-body code GRAINS \citep{ferrari2017,ferrari2020}. The main benefit of using GRAINS is that it is based on the open-source C++ library CHRONO::ENGINE \citep[CHRONO,][]{Chrono2016}, which enables the use of irregularly shaped particles more true to the real shape of boulders in rubble-pile aggregates. As a result, GRAINS can capture key granular mechanics such as particle-particle interlocking, polyhedral contacts and off-centre collisions, which are diminished when using spherical elements. We have recently made significant upgrades to GRAINS to improve its contact model, gravitational force calculations, and input/output operations, outlined in Appendix~\ref{appendix:grains_changes}. The stability of a rubble pile modelled in GRAINS will be largely dependent on the global contact model and its material properties, which in turn govern the level of energy dissipation during contacts between its particles. The global contact model can be either smooth (force-based) or non-smooth (impulse-based), comparable to the soft- and hard-sphere method in other DEM codes such as pkdgrav \citep{richardson2000,stadel2001}. The soft method is well-suited for long-term contacts and uses a Hookean, Hertzian or Flores visco-elastic model to compute the normal and tangential forces that arise due to a contact. The Hookean approach is linear, while the Hertzian and Flores models are non-linear. In this work, we use the smooth contact model with the Hertzian implementation. For a material of a given density, its contact behaviour is governed by a set of parameters that alter friction, restitution, cohesion and stiffness. GRAINS supports four different types of friction: static, sliding, rolling and spinning. The standard values used in this study, summarised in Table~\ref{tab:grains_contact_model}, are motivated by laboratory experiments \citep{Chrono2016,sunday2020}. We note that the rolling and spinning friction is only used for spherical elements, with values of 1.05 and 1.3, motivated by experiments using sand of medium hardness \citep{jiang2015}. It remains unclear how the results from these studies can be translated to zero-gravity environments, but there are ongoing experimental campaigns in low-gravity settings to better constrain our contact model for granular material in space \citep[e.g.][]{vaghi2025}.

Gravitational forces are evaluated using an octree Barnes-Hut implementation with GPU acceleration from CUDA \citep[see Sect.~2.1 in][]{ferrari2020}. We note that GRAINS has previously only been used to explore asteroid-scale problems \citep[e.g.][]{ferrari&tanga2022,agrusa2022,wimarsson2024,wimarsson2025,fodde2026}. We have now extended its use to planetary-scale encounters to model tidal disruptions in this work (see Appendix~\ref{appendix:grains_changes}).

\begin{table}[]
    \centering
    \caption{Material properties in the smooth DEM contact model.}
    \begin{tabular}{l|c|c}
    \hline\hline
          Shape & Irregular & Spheres \\
          \hline 
          Young's modulus (MPa) & 200 & 200 \\
          Poisson's ratio & 0.3 & 0.3 \\ 
         Static friction coefficient & 0.6 & 0.6 \\ 
         Dynamic friction coefficient & 0.6 & 0.6 \\
         Rolling friction coefficient & N/A & 1.05 \\
         Spinning friction coefficient & N/A & 1.3 \\ 
         Cohesive force (N) & 0 & 0 \\ 
         Coefficient of restitution & 0.4 & 0.4 \\ 
         \hline 
    \end{tabular}
    \label{tab:grains_contact_model}
\end{table}

\subsection{Progenitor generation}\label{section:method_aggregate_generation}

\begin{figure}
    \centering
    \resizebox{0.6\hsize}{!}{\includegraphics[width=0.5\linewidth]{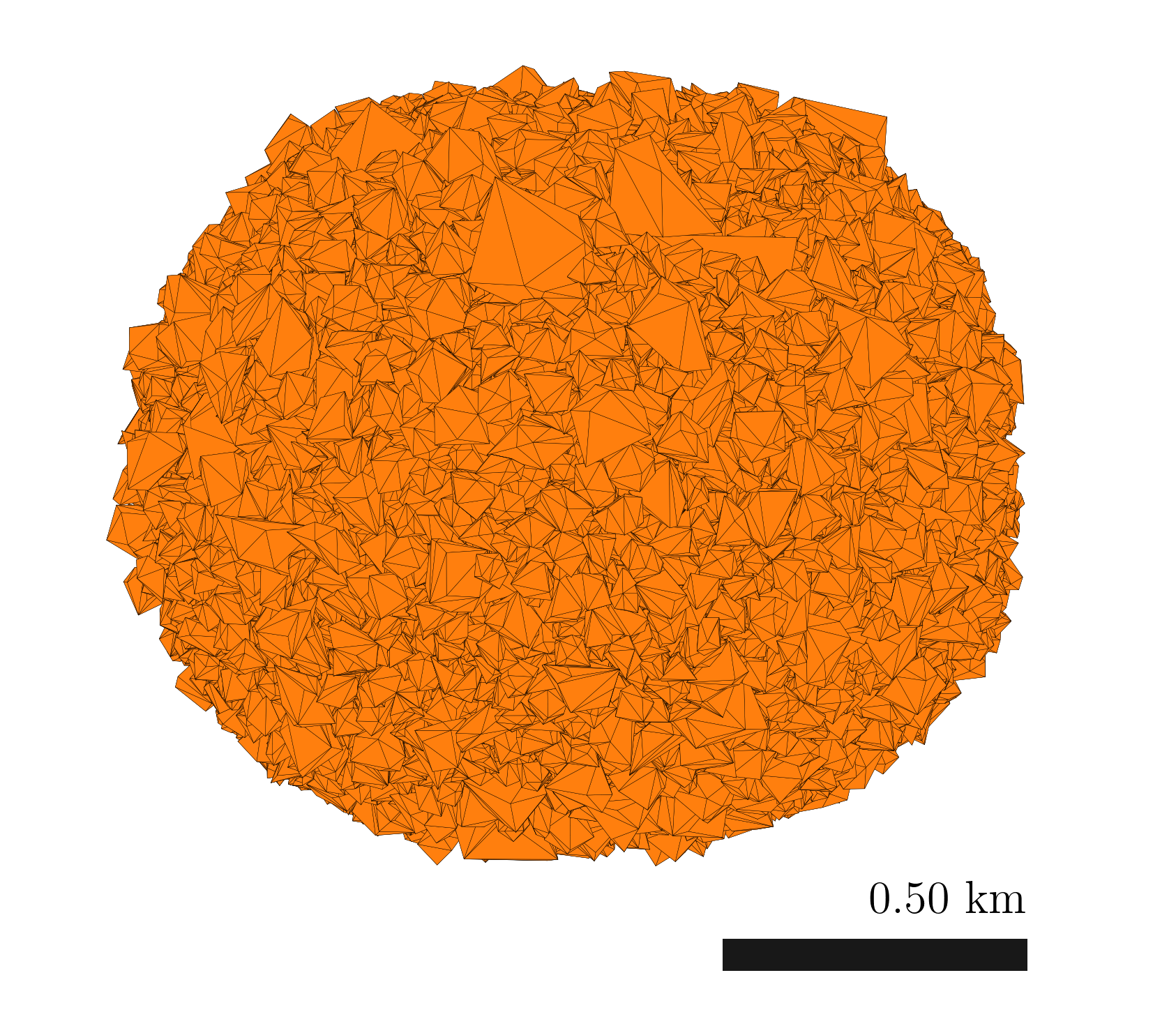}}
    \caption{Standard rubble pile model used in our simulations after its initial aggregation and 50 h of settling in a spinning state. The aggregate consists of 10\,000 irregularly shaped particles in the form of convex hulls.}
    \label{fig:rubble_pile}
\end{figure}

When generating an aggregate with GRAINS, we populated a three-dimensional box with randomly distributed particles that followed some size frequency distribution and let the particles gravitationally accumulate in two stages. The first stage used a linear Hookean contact model with the coefficient of restitution (COR) set to zero and a time step of 0.1 s to quickly accumulate bodies with inelastic contacts. For the fiducial scenario, we used a particle number of $N_p = 10\,000$, and for the high-resolution cases we let $N_p = 50\,000$. At the point where an aggregate had begun to form, we switched over to the second stage that used our standard non-linear Hertzian contacts with a COR of 0.4 and let the body settle with a time step of 0.01 s for a few thousand seconds. For each settling simulation, the transition was set to occur when the number of contacts was significantly larger than the total number of particles, here chosen empirically as $2.5N_p$. For the irregularly shaped particles, we generated a standard number of 16 vertices in a square box of size $D_p$ and computed the corresponding convex hull. The diameter of each box or sphere was selected using the normalised SFD

\begin{equation}\label{equation:particle_sfd}
    f_{D_p}(D_p,D_\mathrm{min},D_\mathrm{mean}) = \lambda_D e^{-(D_p-D_\mathrm{min})\lambda_D},
\end{equation}
\noindent
where $\lambda_D = 1/(D_\mathrm{mean}-D_\mathrm{min})$. For the irregularly shaped particles, there is a natural deviation between the drawn value and the actual diameter of the object due to the randomised vertex generation. Hence, even monodisperse distributions of irregular particles had slight variances in sizes, which was not the case for aggregates consisting of spherical particles. After the initial aggregation and settling concluded, we allowed the rubble pile to settle for an additional 50 h in a spinning state with an angular frequency of $10^{-4}$ rad/s, similar to our target spin rate, using the standard time step of 0.5 s. The settled fiducial rubble pile model, consisting of irregularly shaped particles with a polydisperse SFD, is shown in Fig.~\ref{fig:rubble_pile}, viewed perpendicular to the inertial $z$-axis.

To keep discussion about the different progenitors clear and concise in the text, we henceforth classify their SFDs via the abbreviated forms `poly' (polydisperse) and `mono' (monodisperse), while `high-res poly' refers to high-resolution polydisperse. The particle shapes are simply `irregular' or `sphere', meaning, for example, that our fiducial progenitor is known as `poly irregular' from now on. Here, we also note that while the diameters of the particles in the mono irregular progenitor are highly similar, their shapes might differ due to the random nature of the vertex generation. We provide a summary of properties such as bulk density, resulting SFD parameters, aggregate size and porosity for the models used in this study in Table~\ref{tab:aggregates}. The geometrical properties of each progenitor have been evaluated using the $\alpha$-wrap library of CGAL \citep[the computational geometry algorithms library,][]{cgal:alpha_wrap_3}. This process involves wrapping a three-dimensional `sheet' around the shape points of all constituents of a given aggregate using Delaunay triangulation and shrinking it via subdivision. For the convex hulls, these points are their vertices, while the spheres are represented by ten randomly drawn points from their surface. The resolution of each wrap is determined by the $\alpha$ and offset values, the latter here denoted by $\gamma$. The parameter $\alpha$ determines whether a Delaunay facet is traversable during the shrinking of the mesh by comparing its value to the circumradius of the facet. Hence, a smaller $\alpha$ allows the shrinking process to go deeper into cavities between the mesh vertices, which leads to a finer, more complex mesh. On the other hand, the offset sets the distance between the generated mesh and the underlying isosurface, meaning it determines how tightly the mesh is wrapped around the vertices. The true values of $\alpha$ and $\gamma$ are calculated using the length of the diagonal of the bounding box for the vertices, $d_\mathrm{bbox}$, and two input parameters setting the relative value: $\alpha_\mathrm{rel}$ and $\gamma_\mathrm{rel}$, such that $\alpha = d_\mathrm{bbox} / \alpha_\mathrm{rel}$ and $\gamma = d_\mathrm{bbox} / \gamma_\mathrm{rel}$. We found that values of $\alpha_\mathrm{rel}=20$ and $\gamma_\mathrm{rel}=300$ provided good fits for the number and spatial density of the vertices in our aggregates.

As the point where a body undergoes tidal disruption is strongly related to the bulk density ratio $\rho_M/\rho_m$ (see Sect.~\ref{section:previous_work}) and independent of aggregate size \citep{li2021}, we scale $\rho_\mathrm{bulk}$ for each progenitor to match that of the poly irregular progenitor by altering the material density of its constituents. As a result, the other aggregates do not have the same total mass or volume as the fiducial model. This has negligible effects on the characteristic hyperbolic encounter time, as the mass differences between the aggregates are tiny compared to the mass of the planet.

\begin{table*}[]
    \centering
    \caption{Properties of the different progenitor aggregates used when simulating Earth fly-bys.}
    \begin{tabular}{l|c|c|c|c|c|c|c|c|c}
    \hline\hline 
          Name & $N_p$ & $M_\mathrm{agg}$ & $\rho_\mathrm{bulk}$ & $\rho_\mathrm{p}$ & $D_\mathrm{min}$ & $D_\mathrm{mean}$ & $D_\mathrm{max}$ & $D_\mathrm{agg}$ & Porosity \\ 
         &  & [$10^{12}$ kg] & [g/cm$^3$] & [g/cm$^3$] & [m] & [m] & [m] & [km] &  \\\hline
         Polydisperse irregular$^\star$ & 10\,000 & 2.82 & 1.83 & 3.00 & 42.3 & 62.0 & 217.5 & 1.52 & 0.38 \\
         Polydisperse irregular alt & 10\,000 & 2.78 & 1.83 & 2.93 & 41.6 & 61.9 & 221.3 & 1.42 & 0.38 \\
         Monodisperse irregular & 10\,000 & 6.74 & 1.83 & 2.94 & 69.2 & 88.0 & 88.0 & 1.88 & 0.38 \\
         High-resolution irregular &  50\,000 & 9.51 & 1.83 & 2.83 & 33.4 & 54.1 & 185.6 & 2.16 & 0.35 \\
         Polydisperse sphere & 10\,000 & 16.6 & 1.83 & 2.68 & 60.0 & 91.7 & 469.7 & 2.73 & 0.32 \\
         Monodisperse sphere & 10\,000 & 15.0 & 1.83 & 2.87 & 100.0 & 100.0 & 100.0 & 2.53 & 0.36 \\ \hline 
    \end{tabular}
    \tablefoot{Number of particle constituents, bulk and material densities, min, max and mean diameter of the particles, diameter of the aggregate, as well as its porosity. The fiducial progenitor, used for our standard simulation scenarios, has been marked with $\star$.}
    \label{tab:aggregates}
\end{table*}

\subsection{Fragment detection and tracking}\label{section:method_clusterfinder}

The detection of clustered particles was done with an updated version of the friends-of-friends post-processing method introduced in \citet{wimarsson2024}. Contrary to the previous implementation, the program now tracks fragments through time by matching particle constituents between subsequent frames. First, a given aggregate must contain more than the specified minimum number of particles to be detected, set to ten by default. Here, for two nearby particles $i$ and $j$, we use the particle diameters to determine the linking length, set by $\max(D_i,D_j)$. Second, a detected remnant is assigned a global identifier (global ID) that is then compared with all members of the previous simulation frame. If the new fragment contains at least half the constituents of one of the previous fragments, it is considered a match. The physical properties such as position, velocity, and spin are computed with respect to a fragment's centre-of-mass, with the total moments of inertia being evaluated using the parallel axis theorem. The physical extent of a model is based on the resulting shape from the $\alpha$-wrap method described in Sect.~\ref{section:method_aggregate_generation}, where the diameter is the longest side of the bounding box of the mesh.

Thanks to the global ID system, we can now also track the fate of each fragment when it has been lost, i.e.~when it no longer contains half or more of its previous particle members. If a majority of the lost constituents end up in another fragment that also existed in the previous frame, we classify the event as a merger. If the fragment has not been accreted, it is considered disrupted. Any remnant containing less than half of the previous constituents is given a new, unique global ID. 

\subsection{Simulation setup}\label{section:method_simulation}

\begin{figure}
    \centering
    \resizebox{\hsize}{!}{\includegraphics{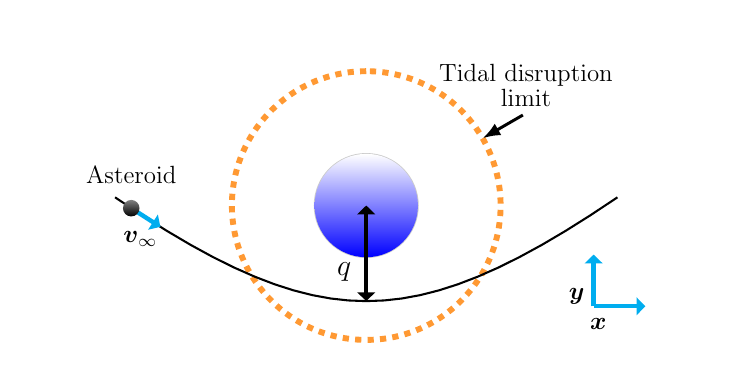}}
    \caption{Diagram of a hyperbolic encounter between an asteroid and a planet (not to scale). The orbit is defined by the encounter velocity at infinite separation, $v_\infty$, and the closest separation, $q$.}
    \label{fig:encounter_orbit}
\end{figure}

\begin{figure}
    \centering
    \resizebox{\hsize}{!}{\includegraphics{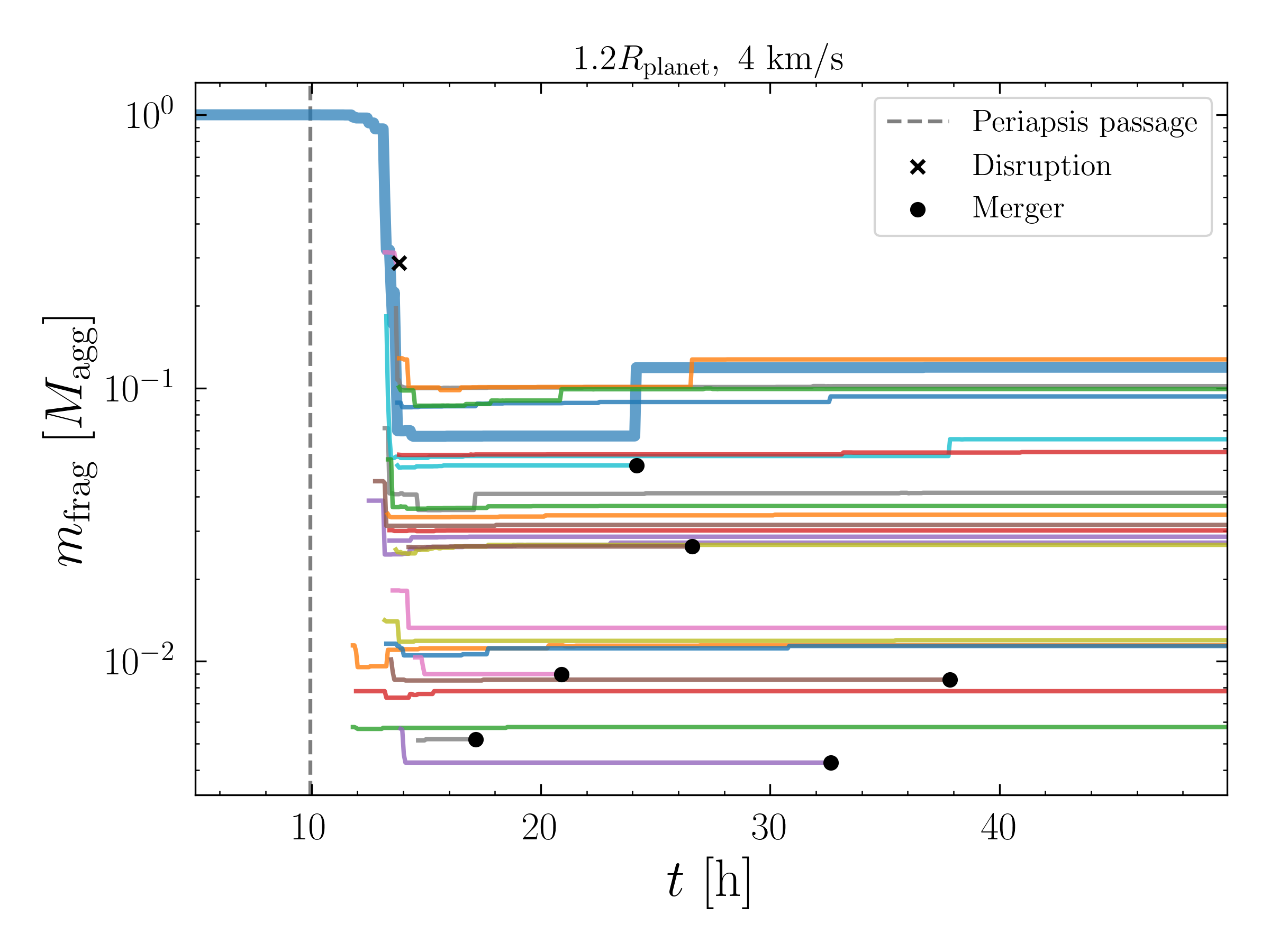}}
    \caption{Mass as a function of time for fragments born in the disruption of the poly irregular progenitor when $q=1.2R_\mathrm{planet}$ and $v_\infty = 4$ km/s. The thicker line shows the mass of the initial aggregate while the other lines represent remnants.}
    \label{fig:mass_vs_time}
\end{figure}

\begin{figure*}[ht]
    \centering
    \resizebox{\hsize}{!}{\includegraphics{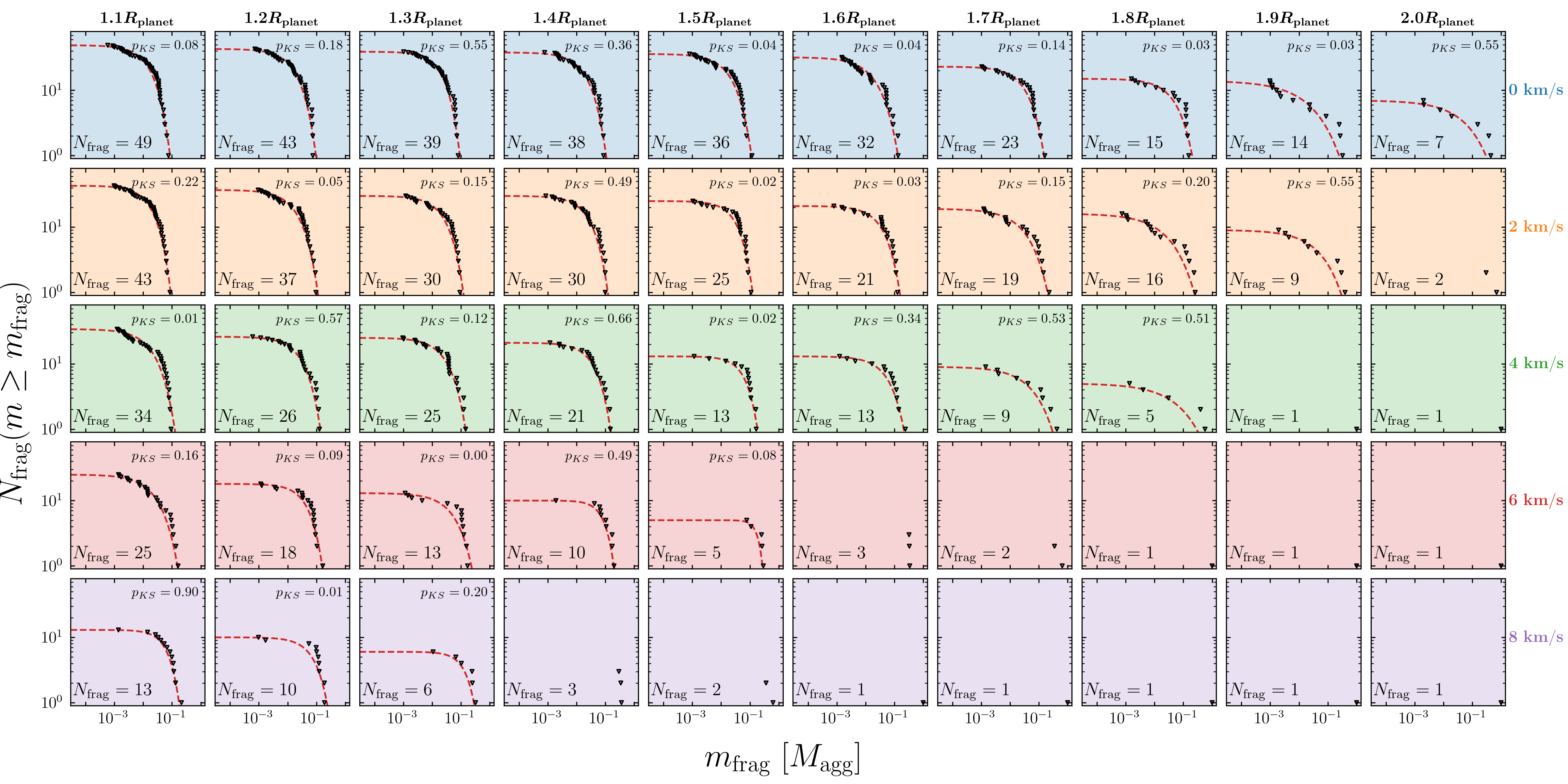}}
    \caption{Distribution of post-disruption masses for the polydisperse irregular aggregate in Fig.~\ref{fig:rubble_pile} given different combinations of heading velocities, $v_\infty$, and periapse distances, $q$. Cases with more than three fragments were also fitted with a Weibull survival function in red. The corresponding p-value from the bootstrapped KS-test for each fit is provided as $p_{KS}$.}
    \label{fig:final_masses_agg1}
\end{figure*}

A tidal encounter was modelled by putting an aggregate onto a parabolic or hyperbolic orbit, where it approaches a given planet in the equatorial plane. Each encounter was simulated in the inert frame of the planet, assumed to have the same properties as Earth ($R_\mathrm{planet}=6378$ km, $\rho_M = 5.51$ g/cm$^{3}$). To compute the orbits, we solved the two-dimensional two-body problem such that we could define each orbit by two parameters: the encounter velocity at infinite separation, $v_\infty$, and the distance at closest approach, $q$. A visualisation of the initial conditions is shown in Fig.~\ref{fig:encounter_orbit}. To ensure that the rubble pile was outside the strong tidal influence of the Earth at the onset of the simulation, we chose a distance between the fluid and rigid Roche limits such that $d_\mathrm{limit} = 2(\rho_M/\rho_m)^{1/3}R_\mathrm{planet}$ and placed the aggregate at $10d_\mathrm{limit}$ from the planet. For the spin, we chose an initial prograde rotational period of 4.3 h based on the prescription in \citet{zhang2020b}. We then simulated each encounter for 72 h employing the minimum residual solver method in CHRONO with a tolerance of $10^{-5}$ and a maximum of 300 iterations. Each simulation used a time step of 0.5 s, chosen to properly resolve the characteristic Hookean contact times for the minimum diameter particles in our systems \citep[see Eqs.~(27)-(29) and Appendix A in][]{sunday2020}.

\section{Tidal disruption simulations}\label{section:results_fiducial}

Our initial set of simulations was performed using the fiducial aggregate in Fig.~\ref{fig:rubble_pile}, consisting of irregularly shaped particles with a polydisperse SFD. The data set contains simulations with periapsis distances $q/R_\mathrm{planet}$ between 1.1 and 2.2 in increments of 0.1 and encounter velocities of 0, 2, 4, 6 and 8 km/s. For an Earth-like escape velocity, $v_\mathrm{esc}$, of 11.18 km/s, these values correspond to $v_\infty/v_\mathrm{esc}$ of 0, 0.18, 0.36, 0.54 and 0.72 at the planetary surface\footnote{To evaluate the $v_\infty/v_\mathrm{esc}$ ratio at different periapsis distances, simply multiply these values by $(q/R_\mathrm{planet})^{-1/2}$.}. In Fig.~\ref{fig:mass_vs_time}, we show an example of the evolution of mass over time for the case $q=1.2R_\mathrm{planet}$ and $v_\infty = 4$ km/s from five hours before periapsis passage to 40 hours after. Each track in the plot represents a unique fragment, with markers indicating the nature of its end, whether it disrupted (cross) or merged with a more massive companion (circle) per the prescription in Sect.~\ref{section:method_clusterfinder}. Here, we only plot the trajectory of fragments with masses larger than $0.005M_\mathrm{agg}$ that survive for more than 2500 time steps. The mass stabilises for all fragments from hour 40 onwards, as most remnants become sufficiently separated to not transfer more mass to each other. That being said, some fragments remain bound in highly eccentric binary systems, similar to what is found by \citet{walsh2006}, and might merge or exchange mass at a later point in their dynamic evolution. We opt not to investigate the nature of these systems in this work and will return to the topic of binary formation in a follow-up study. In any case, the length of our simulations is evidently sufficient to properly resolve the initial mergers and disruptions of fragments, even for parabolic encounters. 

First, we use the parabolic orbit as a reference to constrain that the progenitor undergoes catastrophic disruption for periapsis distances smaller than 2.1$R_\mathrm{planet}$. Beyond this point, the progenitor retains most of its mass and often only experiences reshaping in the form of elongation. We plot the distributions of masses for all cases where $q < 2.1R_\mathrm{planet}$ in Fig.~\ref{fig:final_masses_agg1}, also showing the number of fragments, $N_\mathrm{frag}$. For each $(v_\infty, q)$ combination that produced more than three fragments, we also fit a cumulative Weibull distribution to the data using the stats module of scipy \citep{scipy}. The quality of each fit is indicated with the p-value determined from a bootstrapped Kolmogorov-Smirnov (KS) test, $p_{KS}$. We describe the details of this method in Appendix~\ref{appendix:weibull} and provide all the KS test statistics and Weibull parameters in Table~\ref{tab:weibull_fits}. From the results, we find that most mass distributions appear Weibull in nature, both visually and statistically, with 28 out of 38 samples having $p_{KS}>0.05$. Especially the fits for periapse distances between 1.1 and 1.4$R_\mathrm{planet}$ with a low encounter velocity are of high quality. For the cases further out from the planet where we still see disruptions, such as $(v_\infty,q) = (0\ \mathrm{km/s},1.9R_\mathrm{planet})$ or $(2\ \mathrm{km/s},1.6R_\mathrm{planet})$, we cannot yet conclude if the fits are of lower quality because of a lack of data points, or if the divergences originate from the physics involved. We return to this topic in Sect.~\ref{section:results_shape_study}.

Evaluating the general trends of disruption in our sample, we observe that catastrophic S-class disruptions (largest remnant mass less than $0.5M_\mathrm{agg}$) occur as far out as 2.0$R_\mathrm{planet}$ for parabolic encounters, while this limit moves further and further in with decreasing encounter time. The respective periapsis distances for which we get S-class disruptions for 2, 4, 6 and 8 km/s are 1.9, 1.7, 1.6 and 1.4$R_\mathrm{planet}$. While direct comparison with other studies remains outside the scope of our study, the poly irregular progenitor is notably more easily disrupted than the aggregate consisting of monodisperse, pseudo-irregular particles used in \citet{zhang2020b}, even when they use a low friction angle. Though their higher bulk density (2.43 g/cm$^{3}$ leading to a $d_\mathrm{FRL}$ of $3.21R_\mathrm{planet}$ versus our $3.52R_\mathrm{planet}$) will certainly contribute to this difference, it may not tell the entire story given the differences in our model implementations, including contact models and treatment of non-spherical shapes. Our results are more easily related to the around 1.6 g/cm$^{3}$ bulk densities of \citet{marohnic2026}, who investigate the effect of particle shape and number, using the same numerical code as \citet{zhang2020b}. Performing a trial with $v_\infty = 2$ km/s, instead using a non-rotating progenitor for different values of $q$, it effectively retains its initial mass outside $1.6R_\mathrm{planet}$, only losing a maximum of nine particles further out. We find that S-class disruptions occur up to 1.5$R_\mathrm{planet}$, which produces a largest remnant of 0.30$M_\mathrm{agg}$, while the corresponding mass for the B-class disruption of $q=1.6R_\mathrm{planet}$ is 0.52$M_\mathrm{agg}$. This is more analogous to their high-resolution 10\,000 particle scenario for monodisperse, spherical elements with pseudo-polyhedral contacts. A more detailed comparison will require larger data sets and a better understanding of how the various parameters, such as friction, coefficient of restitution, and Young's modulus, as well as how granular mechanics are captured across different implementations of contact models. Instead, we proceed by focusing our efforts on constraining the effects of polydispersity and particle shape within the scope of GRAINS.

\section{Effect of rubble pile heterogeneity} \label{section:results_heterogeneity}

\begin{figure*}
    \centering
    \resizebox{\hsize}{!}{\includegraphics{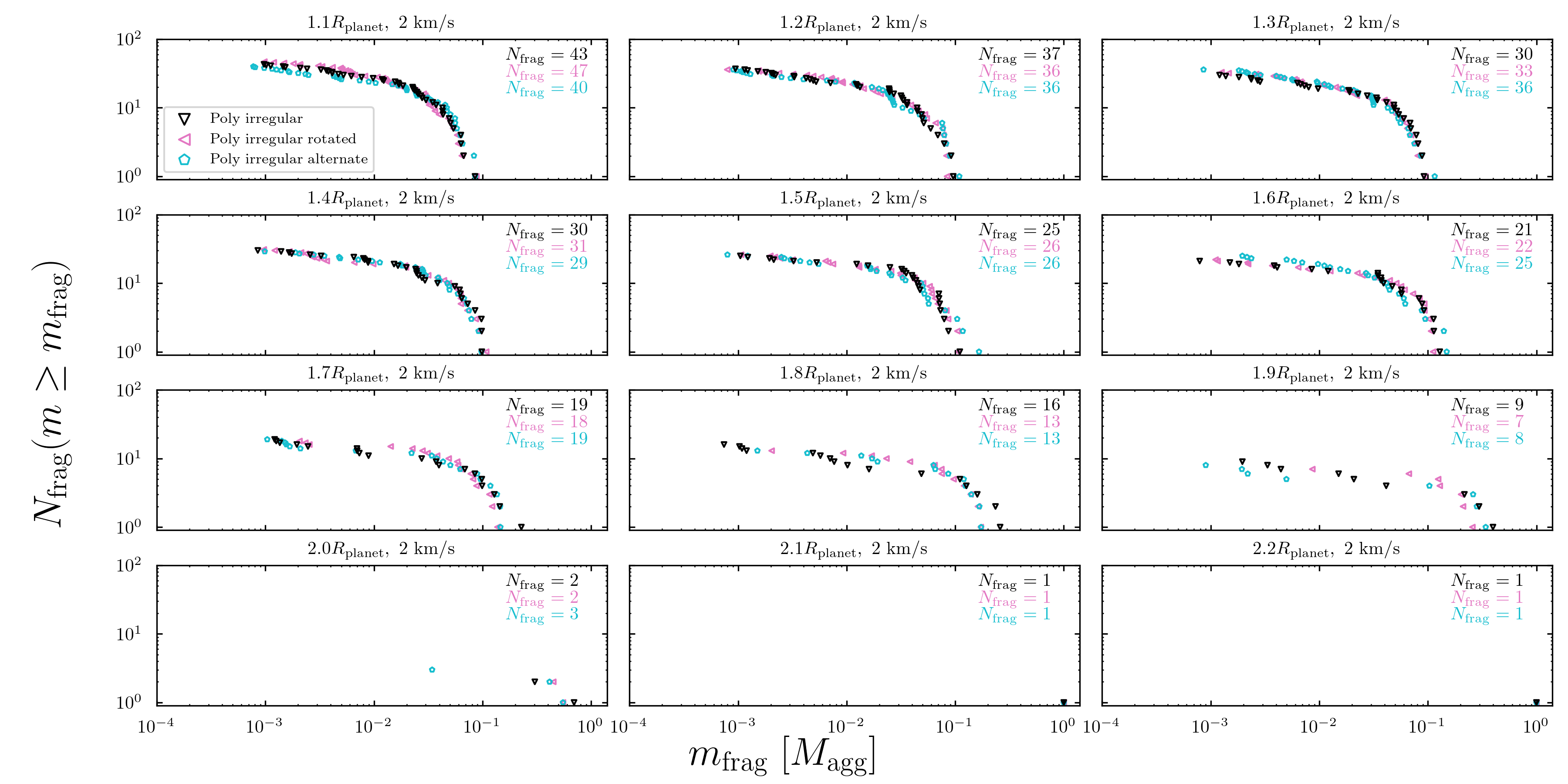}}
    \caption{Distribution of post-disruption fragment masses in terms of total progenitor mass for the fiducial rubble pile and an aggregate with the same particle shape and SFD but a different random seed. The `poly irregular rotated' case refers to the fiducial progenitor being rotated 90$^\circ$ before the start of the simulation. All cases have orbits defined by $v_\infty = 2$ km/s and increasing values of periapse distance, $q$, from top left to bottom right.}
    \label{fig:shape_study_fiducial_alt}
\end{figure*}

Capturing the general effect of heterogeneity is a difficult task given the enormous parameter space for particle properties. Moreover, DEM studies generally struggle with limitations from resolution and the inability to capture the effect of dust in the form of regolith that could, for example, provide cohesion which would considerably alter the dynamics and internal strength of a rubble pile \citep{hirabayashi2015_cohesion,sanchez2014,sanchez2016,zhang2018}. In a first attempt, we performed a number of simulations with rubble piles consisting of different random configurations of particles but the same bulk density, using the different progenitors in Table~\ref{tab:aggregates}. In each set of simulations, the given progenitor undergoes a close encounter with Earth with a hyperbolic orbit defined by $v_\infty = 2$ km/s and the same periapse distances $q/R_\mathrm{planet}$ as in Sec.~\ref{section:results_fiducial}. The fiducial progenitor, poly irregular, was chosen as the basis for comparison with the other aggregates. We find that while the shape of the cumulative distribution of masses is predictable and consistent for specific combinations of particle shape and SFD, the number of remnants generated from a given hyperbolic encounter sees significant variance, even with slight changes in internal structure. 

\subsection{Effect of random packing}\label{section:results_packing}

To highlight this stochastic behaviour, we show the distribution of masses for three different cases consisting of irregular particles with the same SFD in Fig.~\ref{fig:shape_study_fiducial_alt}. The black, upside-down triangles represent the fiducial aggregate, while the pink sideways triangles correspond to the same progenitor, but rotated 90 degrees around its $z$-axis before the onset of the simulation. The cyan pentagons represent a similar aggregate but with a different random seed for drawing initial positions and diameters in the aggregation process from Sect.~\ref{section:method_aggregate_generation}. While the distributions of masses follow similar shapes, clearly also Weibull in nature, there are significant deviations in the mass of the largest remnant and the total number of fragments, especially for periapsis distances between $1.7$ and $2.0R_\mathrm{planet}$. The three distributions ultimately converge at the larger values of $q$, and all progenitors remain intact at $q>2.0R_\mathrm{planet}$. Hence, when simulating tidal disruptions for a randomly packed, heterogeneous rubble pile with a set of given properties, we should expect to get a good indicator for the general distribution of fragment masses for similar progenitors. That being said, since the total fragment number and the mass of the largest remnant fluctuate even for the same progenitor with two different initial geometric rotations, we argue that these parameters should be used with care, especially when estimating how well a numerical model can capture the details of observed tidal disruption events such as Shoemaker-Levy 9.

\subsection{Varying particle shape and SFD}\label{section:results_shape_study}

\begin{figure*}
    \centering
    \resizebox{\hsize}{!}{\includegraphics{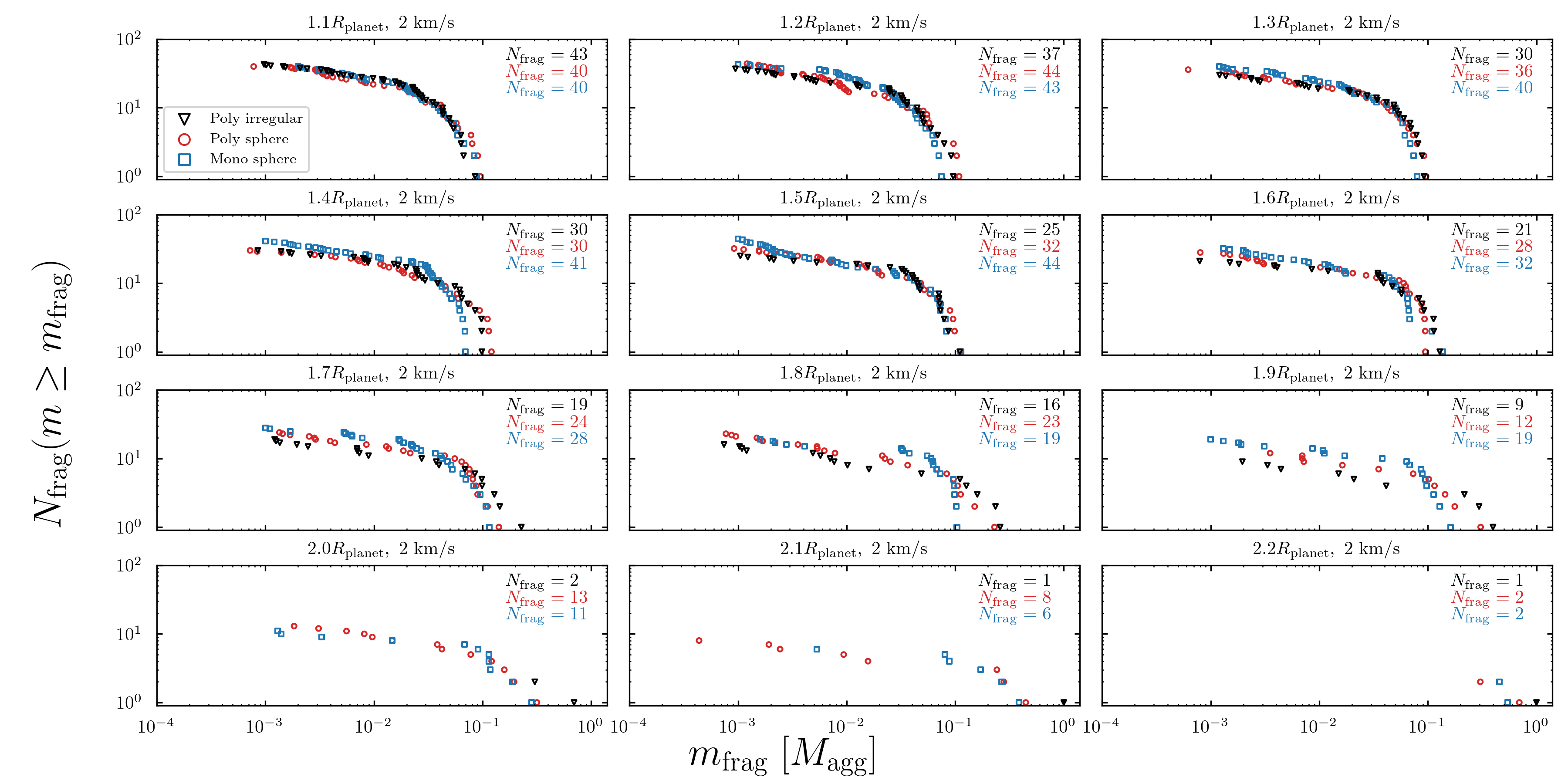}}
    \caption{Same as Fig.~\ref{fig:shape_study_fiducial_alt}, but for comparison with progenitors consisting of spherical particles with a monodisperse or polydisperse SFD.}
    \label{fig:shape_study_masses_spheres}
\end{figure*}

\begin{figure*}
    \centering
    \resizebox{\hsize}{!}{\includegraphics{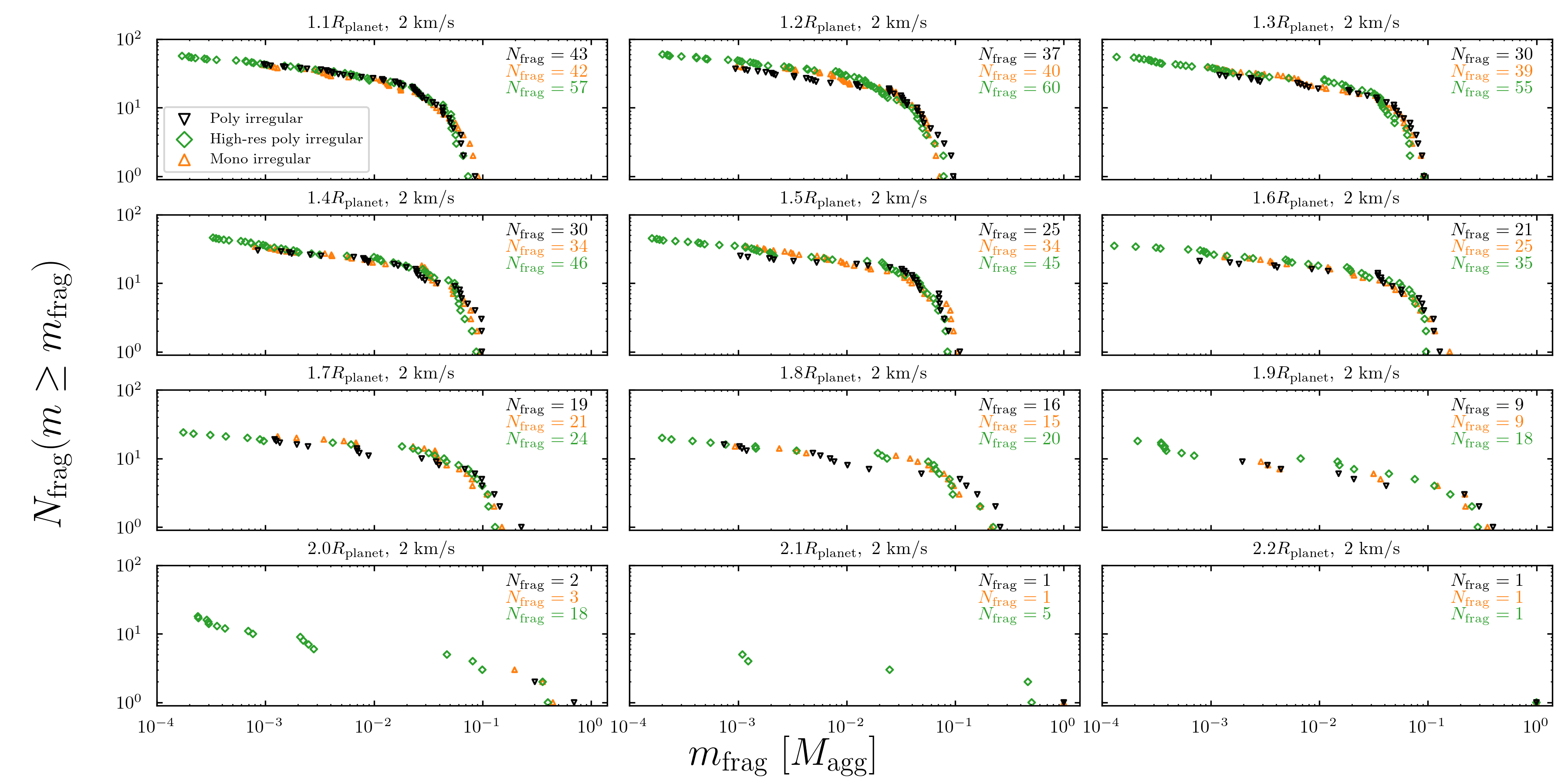}}
    \caption{Same as Fig.~\ref{fig:shape_study_fiducial_alt}, but for irregular particles with a monodisperse or high-resolution polydisperse SFD.}
    \label{fig:shape_study_masses_irregular}
\end{figure*}

Differences in the mass distribution of fragments become more tangible when considering different shapes and SFDs. In Fig.~\ref{fig:shape_study_masses_spheres}, we show the respective distributions for the fiducial, mono sphere (blue circles), and poly sphere (red squares) progenitors. Contrary to the case of Fig.~\ref{fig:shape_study_fiducial_alt}, the shapes of the distributions remain similar only up to the case where $q=1.6R_\mathrm{planet}$. Beyond this point, we see clear divergences in shape that cannot be attributed to random fluctuations determined by particle packing. While the number of fragments is consistently higher when using spherical elements, in line with \citet{marohnic2026}, the mass of the largest remnant changes drastically from $1.7R_\mathrm{planet}$ compared to the fiducial case. Notably, the progenitors with spherical elements also still undergo disruption at $2.1$ and $2.2R_\mathrm{planet}$. This trend indicates that somewhere between a periapsis distance of $1.6$ and $1.7R_\mathrm{planet}$, there is a transition into a new regime where particle shape becomes more relevant. Comparing the substantially different largest remnant masses for the poly and mono sphere cases when $q$ is 1.7 or $1.8R_\mathrm{planet}$, we see that the SFD also strongly affects the resulting mass distribution among fragments. Perhaps unintuitive, the distance at which this change occurs does not coincide with the changeover from S-class to B-class disruption, which for the poly irregular progenitor happens between $1.9$ and $2.0R_\mathrm{planet}$. Given that the classes typically used within the community cannot accurately capture this transition, we suggest adopting the nomenclature of `super-catastrophic' disruption from \citet{granvik2016}, indicating almost complete disintegration. Henceforth, we refer to this as an `SC-class disruption', leading to largest remnant masses, $m_\mathrm{lr}$, less than 20\% of the initial progenitor mass, meaning an S-class disruption follows $0.2M_\mathrm{agg}\leq m_\mathrm{lr} < 0.5M_\mathrm{agg}$. We motivate the SC-class mass criterion at the end of this section.

\begin{figure}
    \centering
    \resizebox{\hsize}{!}{\includegraphics{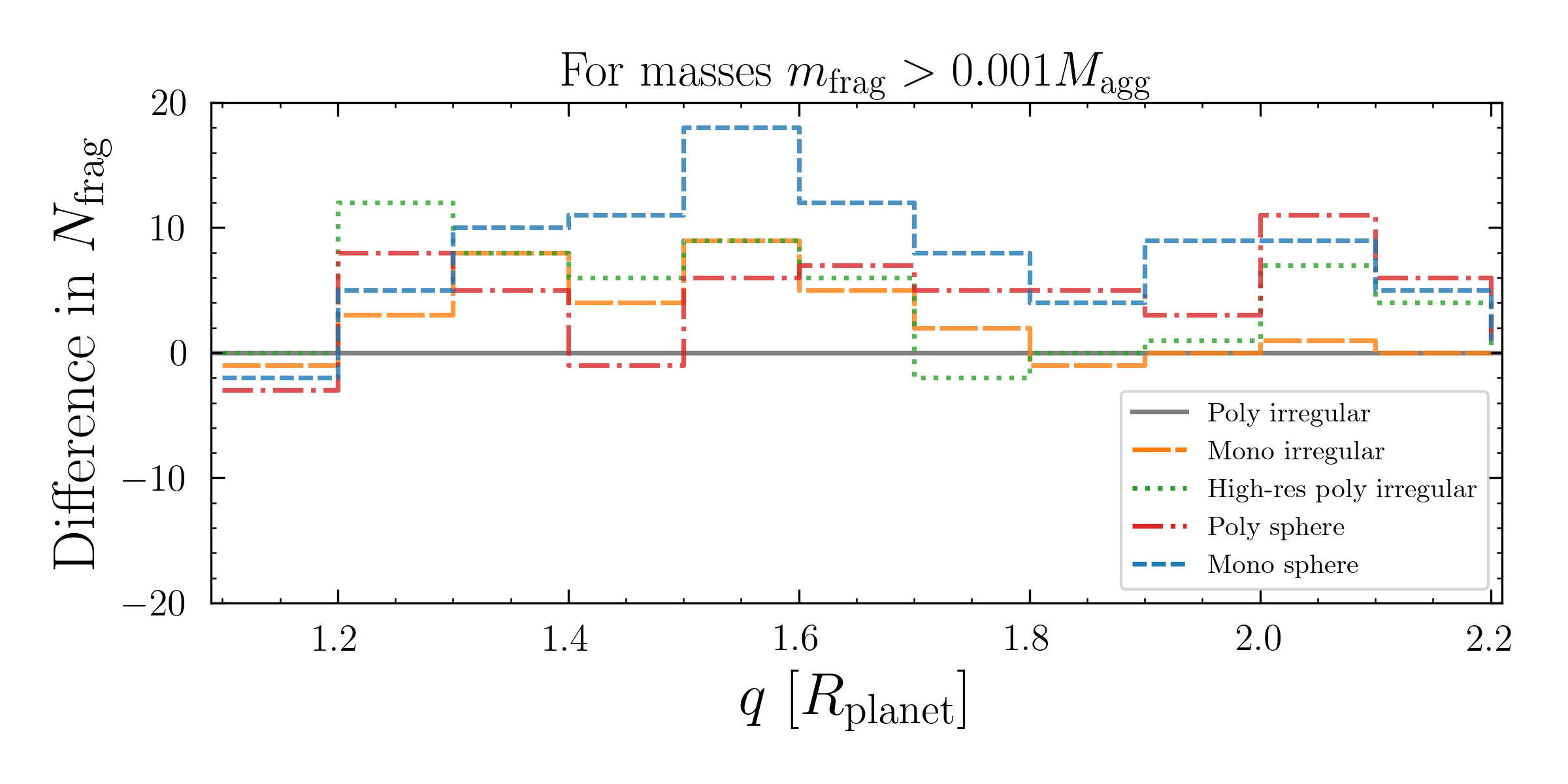}}
    \caption{Difference in fragment number compared to the poly irregular aggregate for various periapse distances, $q$.}
    \label{fig:shape_study_numbers}
\end{figure}

\begin{figure*}
    \centering
    \resizebox{\hsize}{!}{\includegraphics{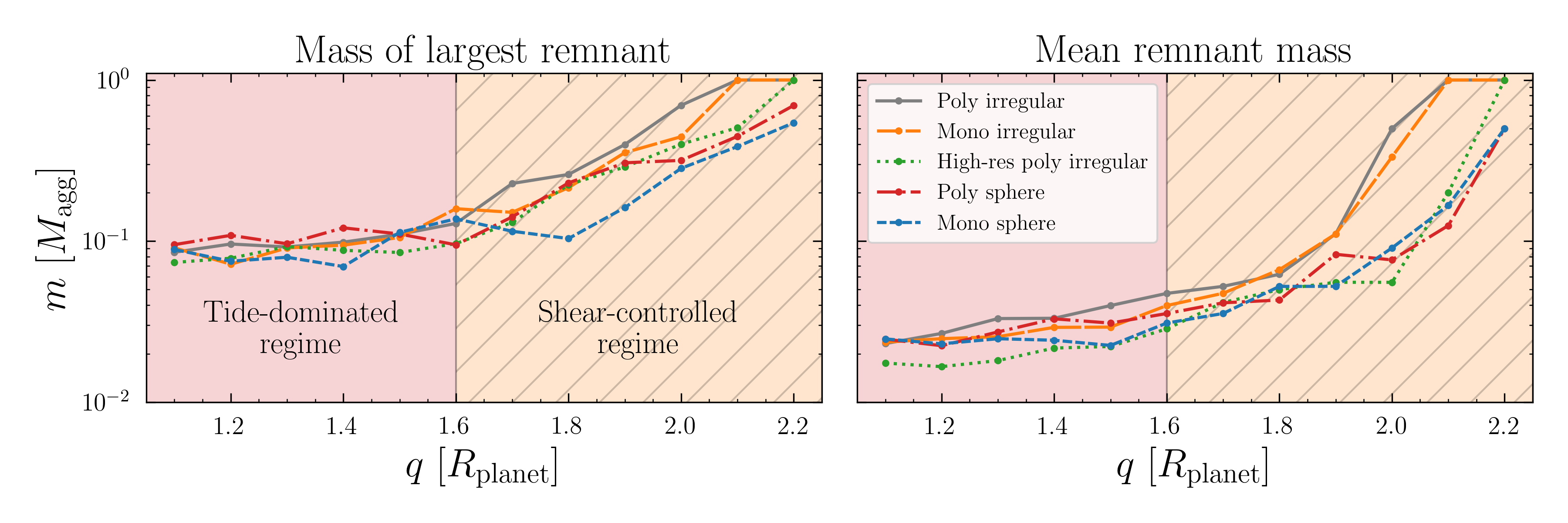}}
    \caption{Mass of the largest remnant (left) and the mean mass of all fragments with $m_\mathrm{frag}>0.001M_\mathrm{agg}$ (right) as a function of periapse distances, $q$, for different progenitors. We show the transition between the tide-dominated (light red) and shear-controlled (orange hatched) regimes for the poly irregular progenitor}
    \label{fig:shape_mass_lr_mean}
\end{figure*}

To better understand the influence of particle shape and SFD, we also look at the distribution of masses for the mono irregular (yellow triangles) and high-res poly irregular (green diamonds) progenitors in Fig.~\ref{fig:shape_study_masses_irregular}. Here, we verify that the shapes of the distributions for the fiducial case are not caused by resolution or the choice of SFD. Even with 50\,000 particles and a steeper polydisperse SFD or monodisperse sizes, we still observe the same Weibull-like pattern up until a periapse distance of $1.7R_\mathrm{planet}$. From this distance to $2.1R_\mathrm{planet}$, there are once more noteworthy discrepancies. Starting with the mono irregular progenitor, the distribution remains less top-heavy than for the corresponding fiducial case at 1.9 and $2.0R_\mathrm{planet}$, leading to similar mean fragment masses but lower masses for the largest remnant. For the high-res poly irregular progenitor, there is a clear divergence in shear resistance, as it produces far more remnants at $2.0R_\mathrm{planet}$ and even disrupts at $2.1R_\mathrm{planet}$ while the other progenitors stay intact. Despite there being a strong variance in fragment number here for the high-res poly case, it is important to keep in mind how we identify fragments, setting a static requirement of at least ten particles. With the steeper SFD, it has smaller average particle masses, which results in lower average fragment masses. This becomes clearer when studying the distributions in Figs.~\ref{fig:shape_study_masses_spheres} and \ref{fig:shape_study_masses_irregular}, focusing on the differences in fragment number compared to the fiducial case when requiring that all fragments must be more massive than $0.001M_\mathrm{agg}$, which is shown in Fig.~\ref{fig:shape_study_numbers}. While the initial differences at $q\leq 1.6R_\mathrm{planet}$ appear mostly random, a pattern emerges beyond this point where we consistently get more fragments for spherical elements. The different SFD cases for the irregular particles end up more similar to the fiducial progenitor, with the high-res poly scenario producing more remnants further out. Nevertheless, it is clear we need a more robust metric for characterising differences in tidal disruption outcomes, based on the analysis in Sect.~\ref{section:results_packing}. We proceed by comparing the masses of the largest remnant (left plot), as well as the mean fragment mass (right plot) in Fig.~\ref{fig:shape_mass_lr_mean}. From these results, the aforementioned patterns become clearer, as we can identify a point of transition to a regime at a periapse distance of around $1.6R_\mathrm{planet}$ where the increased internal strength from granular mechanisms such as interlocking, dilatancy, and polyhedral contacts begins to matter. We define the two regimes as the:

\begin{enumerate}[nosep]
    \item tide-dominated regime: tidal stresses overwhelm internal shear resistance, producing SC-class disruptions;
    \item shear-controlled regime: shear resistance from granular effects such as interlocking becomes relevant and limits bulk motion, producing S-, B- or M-class disruptions.
\end{enumerate}

The transition between these regimes occurs at different values of $q$ for aggregates with spherical elements and/or an SFD more dominated by small particles. If we return to Fig.~\ref{fig:shape_mass_lr_mean} and simply look at the largest remnant mass for the fiducial progenitor at $1.7R_\mathrm{planet}$, it is around $0.23M_\mathrm{agg}$. This value also seems representative for the mono and high-res irregular, and poly sphere cases, which exhibit a transition at $1.8R_\mathrm{planet}$ instead, all with $m_\mathrm{lr}/M_\mathrm{agg}$ equal to 0.21. This observation motivates our choice of mass criterion for SC-class disruptions as $m_\mathrm{lr} < 0.2M_\mathrm{agg}$. However, when we take into account the largest remnant masses of the alternative poly irregular aggregates at $1.7R_\mathrm{planet}$, they exhibit lower values for $m_\mathrm{lr}/M_\mathrm{agg}$ of 0.13 and 0.14 (see Fig.~\ref{fig:shape_study_fiducial_alt}). Hence, we emphasise that while the shear-controlled regime is defined at $q>1.6R_\mathrm{planet}$ based on this empirical analysis, a tidal disruption occurring at these distances does not always lead to a mass of $m_\mathrm{lr} \geq 0.2M_\mathrm{agg}$ because of the stochastic nature of the disruption process. To further motivate our identified transition distance, we show in the next section that shear resistance also affects tidal chain morphology and spin rate. 

In both plots of Fig.~\ref{fig:shape_mass_lr_mean}, we have used the criterion of $m_\mathrm{lr} \geq 0.2M_\mathrm{agg}$ for the poly irregular progenitor to show the transition between the tide-dominated (light red) and the shear-controlled regime (orange hatched). If we apply this requirement to the distributions in Fig.~\ref{fig:final_masses_agg1}, the first $(|v_\infty|,q)$ cases that rest in the shear-controlled regime for a given velocity are: $(0,1.9R_\mathrm{planet})$, $(2,1.7R_\mathrm{planet})$, $(4,1.6R_\mathrm{planet})$, $(6,1.4R_\mathrm{planet)})$ and $(8,1.1R_\mathrm{planet)})$. While we lack enough data to derive an empirical predictor for the transition distance, this will be possible with a larger sample size where we consider additional encounter velocities and periapse distances, also including the density ratio as a parameter. However, at first glance, the identified transition points provide an explanation for the less visually satisfying Weibull fits in the disruption cases further out from the planet in Fig.~\ref{fig:final_masses_agg1}. 

\begin{figure*}
    \centering
    \resizebox{\hsize}{!}{\includegraphics{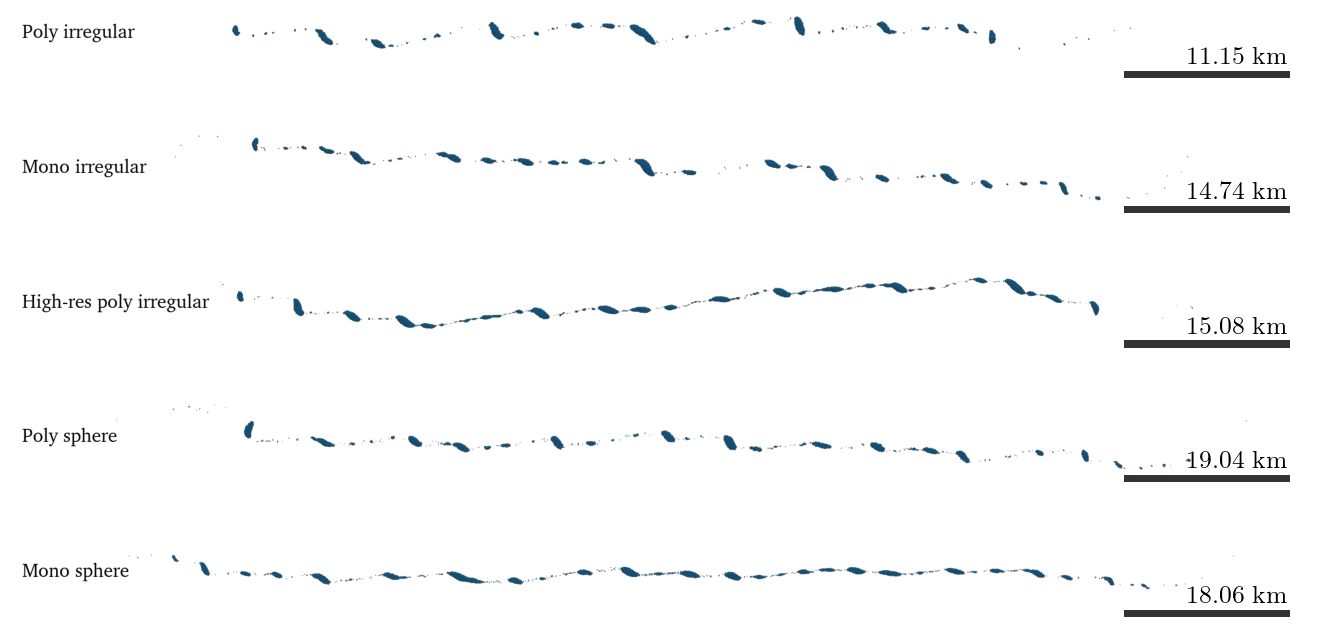}}
    \caption{Tidal chain morphologies for five different progenitors, 7 h after periapsis passage at a distance of $1.7R_\mathrm{planet}$. Each chain has been rotated and aligned with the inertial $x$-axis. Note the different length scales. The associated movies are available online.}
    \label{fig:tidal_chain_lengths_at_365}
\end{figure*}

\subsection{Tidal chain morphology}\label{section:results_tidal_chains}

\begin{figure*}
    \centering
    \resizebox{\hsize}{!}{\includegraphics{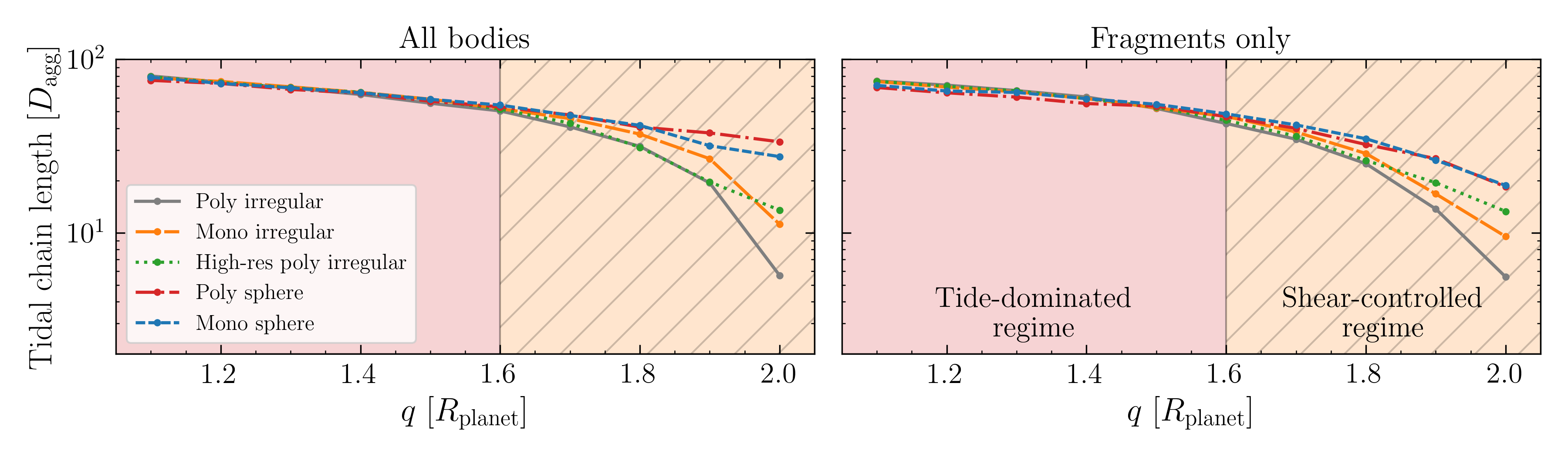}}
    \caption{Tidal chain extent for different progenitors in terms of their initial diameter, $D_\mathrm{agg}$, as a function of periapsis distance $q$ at 7 h after periapsis passage when considering either all particles (left) or only fragments more massive than $0.001M_\mathrm{agg}$ (right).}
    \label{fig:tidal_chain_lengths_vs_q}
\end{figure*}

\begin{figure}
    \centering
    \resizebox{\hsize}{!}{\includegraphics{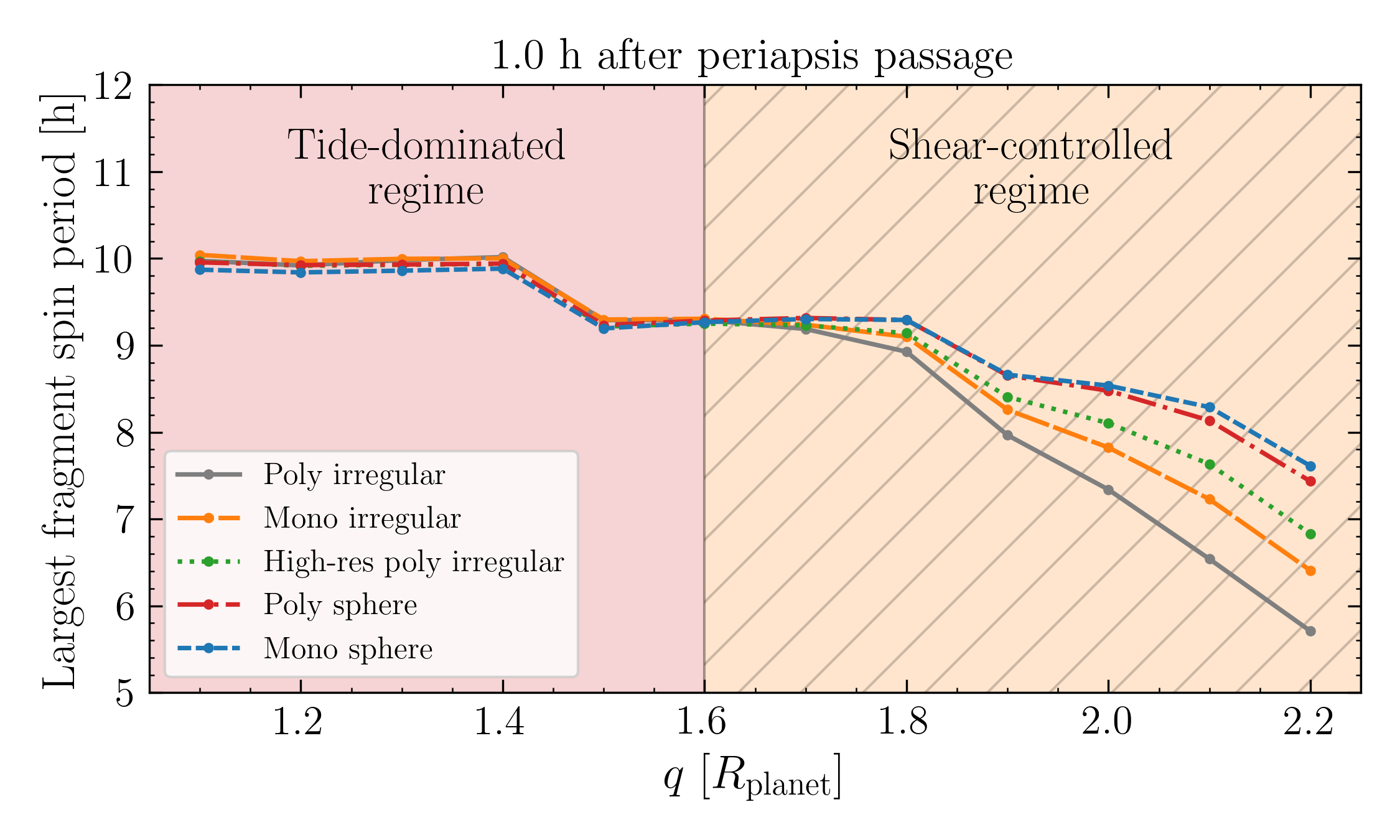}}
    \caption{Spin period of the different progenitors one hour after periapsis passage for increasing values of distance, $q$.}
    \label{fig:largest_fragment_spin_vs_q}
\end{figure}

The crossover to the shear-controlled regime can be inferred more directly from Fig.~\ref{fig:tidal_chain_lengths_at_365}, where we show the tidal chain morphologies for our standard progenitors, 7 hours after periapsis passage when $q=1.7R_\mathrm{planet}$ (for visualisations of these disruptions, see supplementary movies\footnote{The movies are titled \texttt{\{progenitor\_name\}\_1.7Rp\_2kms.mp4} and are shown in the top-down, inert frame of the progenitor, centred on the barycentre of all fragments. The white arrow in the visualisation points towards the planet.}). There is a remarkable divergence in physical extents, where the total span of each chain is strongly affected by particle shape and SFD. While the tidal chain formed from the poly irregular progenitor is only 61 km long, the mono irregular case produces a 90 km long chain, and the corresponding length for the high-res poly progenitor is 93 km. The total lengths of the poly and mono sphere chains are 130 and 121 km, respectively. We note that these values are affected by ejected, free-floating single particles at the edges of the chains, which are difficult to resolve in the image. A comparison between the extents for chain-producing values of $q$, both for all particles and when only considering fragments, is presented in Fig.~\ref{fig:tidal_chain_lengths_vs_q}. In the right-side plot, the poly and mono irregular chains are significantly more similar, with the latter only being a few km longer. On the other hand, the high-res poly progenitor with its steeper SFD shows a similar pattern to the low-resolution case when normalised by its initially larger size, only diverging when disrupted far out from the planet. Considering all bodies, the transition point to the shear-controlled regime emerges distinctly at a periapsis distance of 1.6$R_\mathrm{planet}$. 

A second notable feature of the tidal chains is their spiral patterns. Focusing on the poly and mono sphere cases, even though they have highly similar lengths, the chains show a significant variation in morphology. The poly case has clear overdensities and spiral features, demonstrating how the polydisperse SFD prolongs the disruption duration and gives the progenitor stability. This aligns well with the results from \citet{raducan2024b}, where the authors show how heterogeneous rubble piles with larger boulders emplaced in a continuous medium are less susceptible to disruption by impacts. When we further consider irregularly shaped particles, this effect becomes more pronounced. This behaviour is well-established from previous studies finding that non-spherical particles provide significant structural integrity for rubble piles and make them less susceptible to deformation via tides \citep{movshovitz2012,marohnic2023,marohnic2026,demartini2025}. Hence, it is no surprise that the strongest spiral patterns and smallest extents can indeed be found for the fiducial case, where large boulders, irregular particle shapes and their random packing provide structural integrity from three degrees of heterogeneity.

Given that the progenitors have identical orbits and negligible variations in encounter time, we attempt to quantify the increased resilience against disruption by studying the spin period of the largest fragments 1 h after periapsis passage in Fig.~\ref{fig:largest_fragment_spin_vs_q}. The periods are notably longer (considering the initial value of 4.3 h) for passages closer to the planet, with small variations across the various progenitors, showcasing how the timing of the disruption plays a large role. At this point in the orbit, for values $q<1.6R_\mathrm{planet}$, the disruption is already taking place. Simply calculating the principal axis ratios for the largest remnant in the body-fixed frame shows severe elongation with values over 20. The corresponding structural failure results in long rotational periods. Beyond $1.6R_\mathrm{planet}$, the periods start to diverge, once more highlighting the transition to a more shear-controlled regime. The bodies consisting of irregularly shaped particles withstand disruption for longer, leading to prolonged exposure to the tidal torque and thus shorter rotational periods. Even so, the spin rate does not tell the entire story. For the spherical element progenitors, their periods in Fig.~\ref{fig:largest_fragment_spin_vs_q} remain close to identical at all values of $q$. That being said, when evaluating the mean spin period for the two scenarios, they differ drastically at the corresponding frame shown in Fig.~\ref{fig:tidal_chain_lengths_at_365}, with 14 h for the polydisperse progenitor and 22 h for the monodisperse case. This observation indicates that heterogeneity also plays an important role during the entire disruption phase as fragments form, disrupt, collide and merge. In fact, the two low-resolution polydisperse cases produce considerably higher mean spin rates for their fragments than their monodisperse counterparts throughout most of the simulation. 

\subsection{Stress profiles} \label{section:stress_profiles}

\begin{figure}
    \centering
    \resizebox{\hsize}{!}{\includegraphics{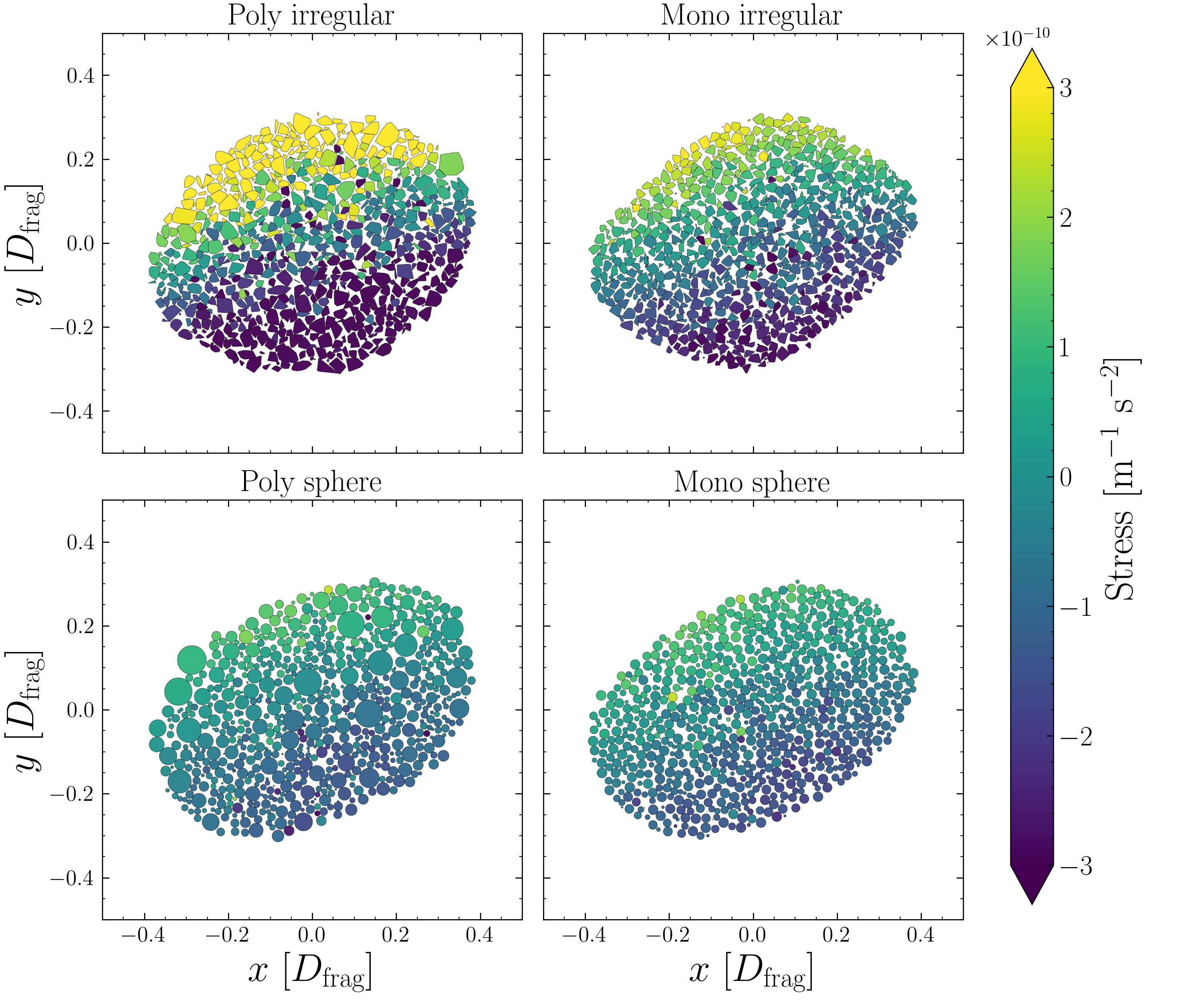}}
    \caption{Per-particle stress slice in the $xy$-plane of four progenitors at periapsis passage when $q=1.9R_\mathrm{planet}$. Each stress value has been computed from the per-particle stress tensor and normalised by the total progenitor mass. The size of each particle may be unrepresentative of the total size, as we only include the area of the intersection between the particle and the slice plane.}
    \label{fig:pressure_profiles}
\end{figure}

\begin{figure}[t]
    \centering
    \resizebox{0.8\hsize}{!}{\includegraphics{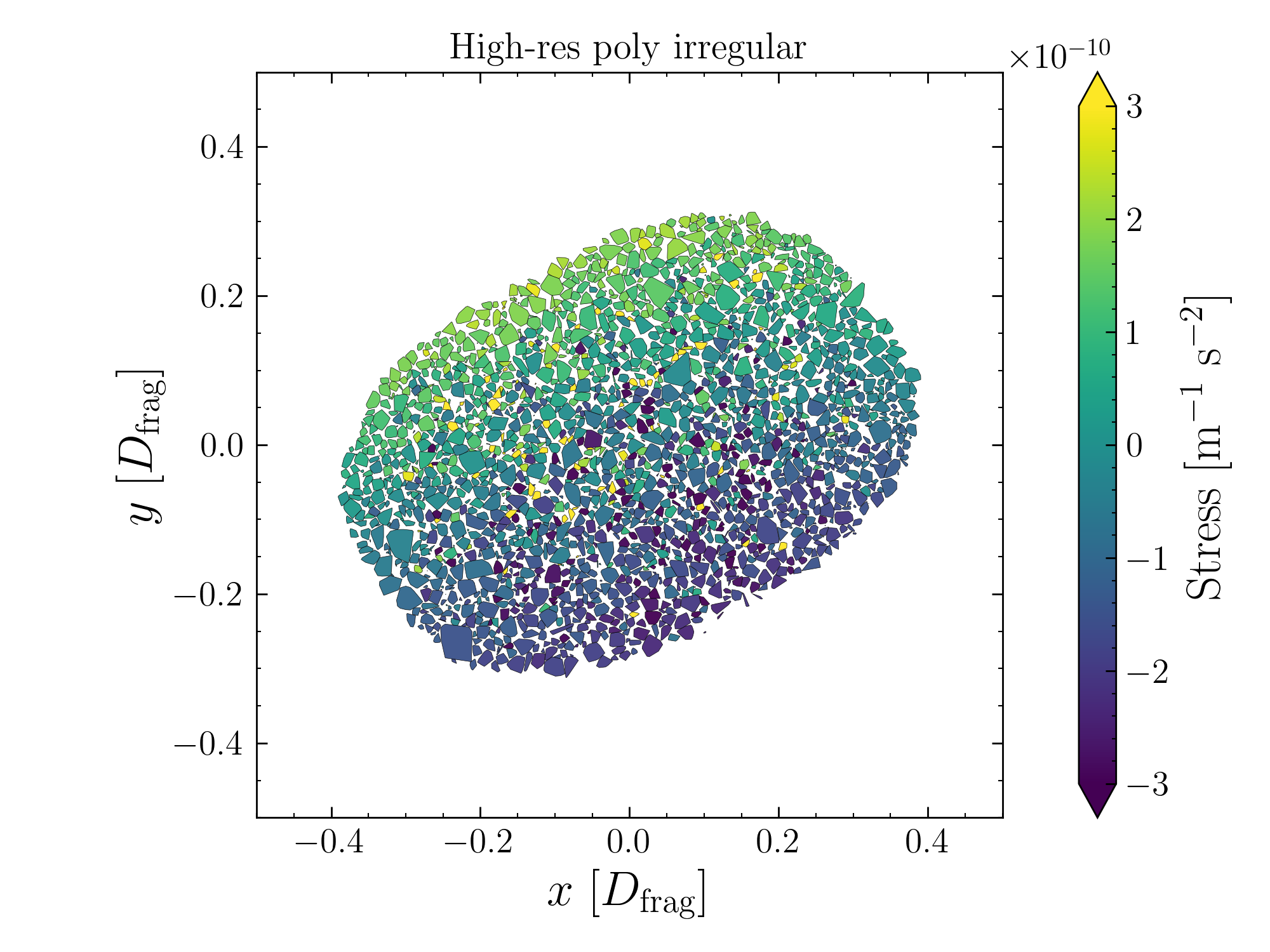}}
    \caption{Same as the panels in Fig.~\ref{fig:pressure_profiles}, but for the high-res poly irregular case.}
    \label{fig:pressure_profile_highres}
\end{figure}

\begin{figure}[t]
    \centering
    \resizebox{\hsize}{!}{\includegraphics{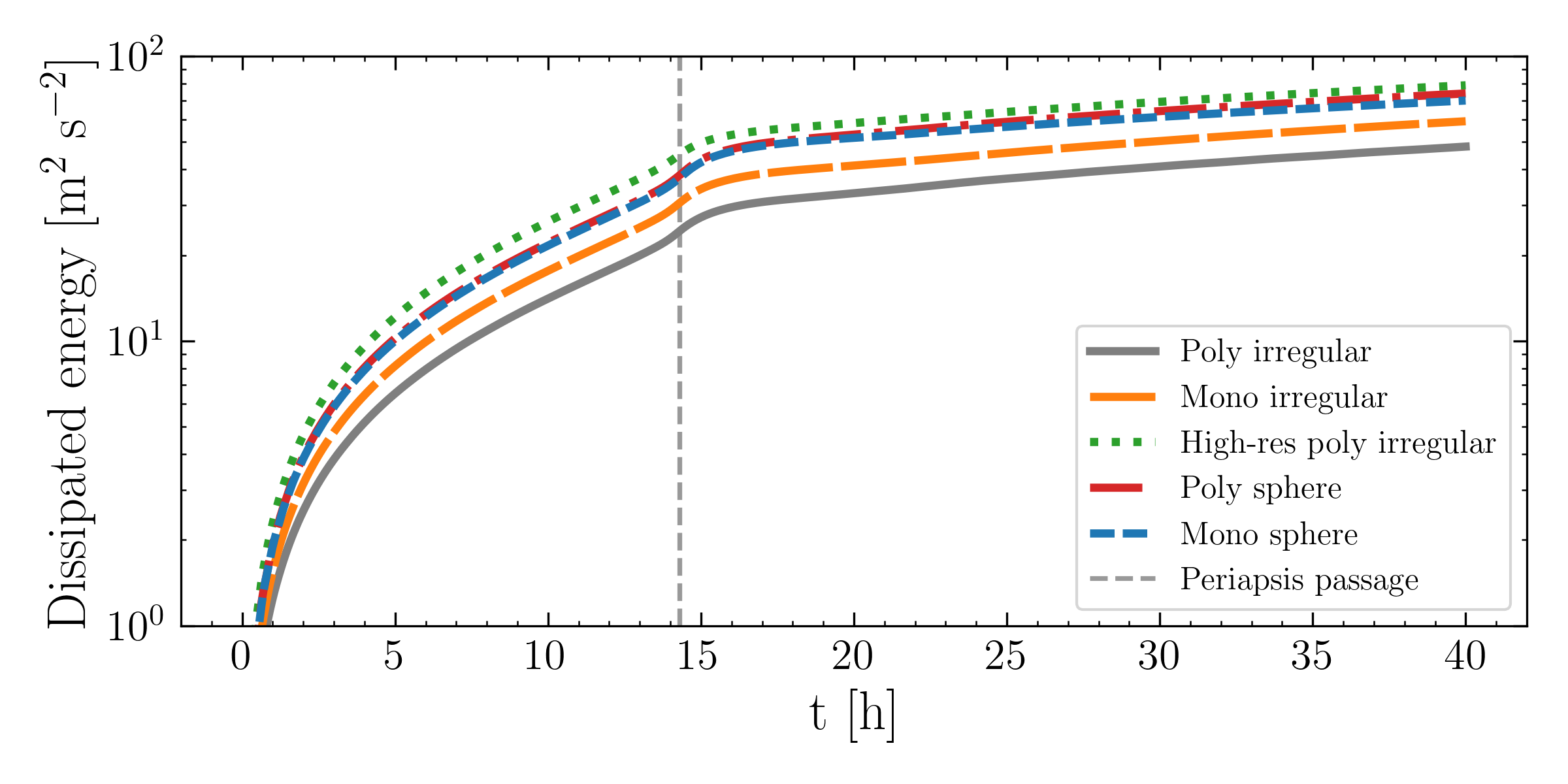}}
    \caption{Cumulative dissipated specific energy from contacts over time for the standard progenitors, normalised by their total initial mass.}
    \label{fig:cumulative_dissipation}
\end{figure}

To characterise the effects of the tidal torque, we analysed the per-particle scalar stress as derived from the stress tensor, following the methodology described in Appendix~\ref{appendix:pressure}. Given the significant overhead from evaluating the stress tensor at each time step, we opted to only perform this analysis for the simulations where $q=1.9R_\mathrm{planet}$, a case that rests firmly in the shear-controlled regime for the fiducial progenitor. In Fig.~\ref{fig:pressure_profiles}, we show two-dimensional slices in the $xy$-plane of our standard low-resolution progenitors at the periapsis passage. The corresponding slice for the high-res poly irregular progenitor can be found in Fig.~\ref{fig:pressure_profile_highres}. The size of each convex hull only represents the area of each intersection between the particle and the $xy$-plane, and the stress values, indicated by the colour of each particle, have been normalised by the total mass of the progenitor. A positive stress indicates compression while a negative value represents decompression. As expected from our previous analysis, the specific stress has substantially higher magnitude for the irregular particles, and even more so for the poly case. Due to the increased internal strength from granular mechanical effects such as interlocking, the individual particles suffer higher stress. The monodisperse progenitor is more easily deformed as it lacks the added effect of polydispersity, which creates `locking points' where individual particles suffer increased compression or decompression, reinforcing the structural integrity of the aggregate. These can be identified as brighter/darker points in predominantly darker/brighter regions or towards the centre. The number of locking points in the slice decreases from top left to bottom right, which also coincides with an increase in deformation. Much like for the analysis in Sect.~\ref{section:results_shape_study} and \ref{section:results_tidal_chains}, the behaviour of the high-res poly irregular case falls in between the other standard progenitors. While it exhibits numerous locking points, the magnitudes of the stress are generally lower, and it undergoes a higher degree of deformation.

The weaker shear strength of the progenitors with spherical elements enables stronger bulk motion, which is also evident in the cumulative energy dissipated into the contact network of our simulations. A comparison between the levels of dissipation for each of the standard progenitor cases can be found in Fig.~\ref{fig:cumulative_dissipation}. As we currently only track the dissipation from contacts using the limited internal bookkeeping in CHRONO per the description in Appendix~\ref{appendix:pressure}, we urge the reader to only infer qualitative patterns from the presented values. While the high-res poly irregular case has the most dissipation across the different progenitors, it can be directly explained by the large increase in total number of contacts that comes with having five times as many particles. More importantly for our discussion, the poly irregular scenario is by far the least dissipative, indicating less internal movement during the tidal encounter compared to its spherical element counterpart and the mono sphere aggregate. It remains to be seen which of the explored SFDs best captures the internal structure of rubble piles, but we can expect better constraints with ESA's Hera \citep{michel2022,michel2025} and Ramses \citep{lazzarin2025} missions that will perform in situ radar observations of rubble piles in the upcoming years \citep{herique2025}.

\section{Conclusion and outlook}\label{section:conclusions}

It is well-established by now that simulating tidal disruptions of rubble piles is a dynamically rich problem with a large parameter space that keeps growing in scope. Yet, previous efforts have significantly narrowed down key mechanisms that govern the nature of a disruption. At this point, it is clear that the bulk density \citep{asphaug1994}, elongation and rotation of the progenitor \citep{richardson1998} influence where a rubble pile disrupts. It is also known that proper modelling of soft contacts in DEM simulations, using realistic particle packing \citep{zhang2020b}, and somehow accounting for the non-sphericity of real boulders and grains, significantly enhances the resilience of a rubble pile against tidal forces \citep{movshovitz2012,zhang2020b,marohnic2023,marohnic2026}. With this numerical study of parabolic and hyperbolic tidal disruptions, we have shown that including a third degree of heterogeneity in the form of polydisperse size frequency distributions (SFDs) can further substantially affect not only the point at which a rubble pile disrupts, but also the nature of its disruption. While unimportant for the most chaotic, tide-dominated region near the planet, we identify a transition to a shear-controlled regime where these heterogeneities become meaningful. Moreover, combining irregular particle shapes with polydispersity, an even higher order of structural reinforcement is reached because of internal `locking points' with local peaks in stress. The resulting structural strength leads to more confined tidal chains, as well as more massive largest remnants. The influence of this effect does, however, decrease when increasing the total number of particles by considering an SFD more dominated by smaller particles, as the surge in particle--particle contacts further enhances flow in our granular media. In turn, the distance at which this shift occurs changes with the internal properties of the aggregate, as well as with the encounter velocity. We identify that said transition distance appears to coincide with a largest remnant mass exceeding or equal to 20\% of the progenitor mass. To distinguish between the two types of catastrophic disruption, we use the nomenclature of \citet{granvik2016} to suggest the introduction of a new `super-catastrophic disruption' class (SC-class) to complement the existing S-, B- and M-classes from \citet{richardson1998}, covering cases where the largest remnant mass is below 20\% of the progenitor mass.

For most progenitors and orbits, the distribution of fragment masses in the tide-dominated regime is well approximated with a Weibull function, but the stochastic nature of the disruptions causes variations in fragment number and largest remnant mass, even for samples from progenitors with highly similar initial conditions. Therefore, we argue that disruption of heterogeneous rubble piles should be treated as both a numerical and statistical problem going forward, especially in the shear-controlled regime. To elaborate, based on our results, to provide an estimate of the bulk density of Shoemaker-Levy 9 from a heterogeneous progenitor (modelling the body as non-rotating with a parabolic orbit), we would have to perform a statistically relevant number of simulations for each bulk density case with different initial geometric rotations. Capturing one outcome will not be sufficient to provide a valid guess. Thankfully, with steady improvements in our computational ability, we can now perform a larger number of simulations in a shorter period of time, with higher resolution than before. Moreover, with high-quality in situ measurements of rubble-pile properties from Hera and Ramses, providing unprecedented estimates of bulk densities, internal structure and surface features, our models will continue to improve and shrink the large parameter space we are currently facing. Combined with additional research into the effect of friction, restitution, and stiffness for our contact models from both numerical and experimental studies, we will be well-equipped to further explore the chaotic nature of tidal disruption and unlock the secrets of small bodies in the Solar System.

\begin{acknowledgements}
J.W. and F.F. acknowledge funding from the Swiss National Science Foundation (SNSF) Ambizione grant No.\,193346. M.J. acknowledges support from SNSF project No.\,200021\_207359. E.F. acknowledges funding from the European Union's Horizon Europe research and innovation programme, grant agreement No. 101077758, ERC TRACES. We thank Robert E. Melikyan for insightful discussion during the course of this study.
\end{acknowledgements}

\bibliographystyle{aa.bst}

\bibliography{References}

@ARTICLE{aggarwal1974,
       author = {{Aggarwal}, H.~R. and {Oberbeck}, V.~R.},
        title = "{Roche Limit of a Solid Body}",
      journal = {\apj},
         year = 1974,
        month = jul,
       volume = {191},
        pages = {577-588},
          doi = {10.1086/152998},
       adsurl = {https://ui.adsabs.harvard.edu/abs/1974ApJ...191..577A}
}

@ARTICLE{agrusa2022,
       author = {{Agrusa}, Harrison F. and {Ferrari}, Fabio and {Zhang}, Yun and {Richardson}, Derek C. and {Michel}, Patrick},
        title = "{Dynamical Evolution of the Didymos-Dimorphos Binary Asteroid as Rubble Piles following the DART Impact}",
      journal = {\psj},
         year = 2022,
        month = jul,
       volume = {3},
       number = {7},
          eid = {158},
        pages = {158},
          doi = {10.3847/PSJ/ac76c1},
archivePrefix = {arXiv},
       eprint = {2207.06995},
 primaryClass = {astro-ph.EP},
       adsurl = {https://ui.adsabs.harvard.edu/abs/2022PSJ.....3..158A}
}

@ARTICLE{agrusa2026,
       author = {{Agrusa}, H. and {Michel}, P.},
        title = "{Tidal disruptions of rubble piles: The case of Phobos}",
      journal = {\aap},
         year = 2026,
        month = feb,
       volume = {706},
          eid = {A353},
        pages = {A353},
          doi = {10.1051/0004-6361/202557988},
archivePrefix = {arXiv},
       eprint = {2602.21912},
 primaryClass = {astro-ph.EP},
       adsurl = {https://ui.adsabs.harvard.edu/abs/2026A&A...706A.353A}
}

@ARTICLE{asphaug1994,
       author = {{Asphaug}, E. and {Benz}, W.},
        title = "{Density of comet Shoemaker-Levy 9 deduced by modelling breakup of the parent 'rubble pile'}",
      journal = {\nat},
         year = 1994,
        month = jul,
       volume = {370},
       number = {6485},
        pages = {120-124},
          doi = {10.1038/370120a0},
       adsurl = {https://ui.adsabs.harvard.edu/abs/1994Natur.370..120A}
}

@article{asphaug1996,
	title = {Size, density, and structure of comet {Shoemaker} - {Levy} 9 inferred from the physics of tidal breakup},
	volume = {121},
	doi = {10.1006/icar.1996.0083},
	number = {2},
	journal = {Icarus},
	author = {{Asphaug}, E. and {Benz}, W.},
	year = {1996},
	note = {Publisher: Academic Press Inc.},
	pages = {225--248},
}

@ARTICLE{binzel2010,
       author = {{Binzel}, Richard P. and {Morbidelli}, Alessandro and {Merouane}, Sihane and {DeMeo}, Francesca E. and {Birlan}, Mirel and {Vernazza}, Pierre and {Thomas}, Cristina A. and {Rivkin}, Andrew S. and {Bus}, Schelte J. and {Tokunaga}, Alan T.},
        title = "{Earth encounters as the origin of fresh surfaces on near-Earth asteroids}",
      journal = {\nat},
         year = 2010,
        month = jan,
       volume = {463},
       number = {7279},
        pages = {331-334},
          doi = {10.1038/nature08709},
       adsurl = {https://ui.adsabs.harvard.edu/abs/2010Natur.463..331B}
}

@article{bottke1999,
	title = {1620 {Geographos} and 433 {Eros}: {Shaped} by {Planetary} {Tides}?},
	volume = {117},
	issn = {1538-3881},
	shorttitle = {1620 {Geographos} and 433 {Eros}},
	url = {https://iopscience.iop.org/article/10.1086/300811/meta},
	doi = {10.1086/300811},
	language = {en},
	number = {4},
	urldate = {2025-11-06},
	journal = {The Astronomical Journal},
	author = {Bottke, W F and Richardson, D. C. and Michel, P. and Love, S. G.},
	month = apr,
	year = {1999},
	note = {Publisher: IOP Publishing},
	pages = {1921},
}

@incollection{burtscher2011,
  title={An efficient CUDA implementation of the tree-based barnes hut n-body algorithm},
  author={Burtscher, Martin and Pingali, Keshav},
  booktitle={GPU computing Gems Emerald edition},
  pages={75--92},
  year={2011},
  publisher={Elsevier}
}

@incollection{cgal:alpha_wrap_3,
  author = {Pierre Alliez and David Cohen-Steiner and Michael Hemmer and C{\'e}dric Portaneri and Mael Rouxel-Labb{\'e}},
  title = {{3D} Alpha Wrapping},
  publisher = {{CGAL Editorial Board}},
  edition = {{5.5}},
  booktitle = {{CGAL} User and Reference Manual},
  url = {https://doc.cgal.org/5.5/Manual/packages.html#PkgAlphaWrap3},
  year = 2022
}

@article{demartini2025,
	title = {The {Influence} of {Internal} {Structure} on {Physical} {Outcomes} of the 2029 {Apophis} {Close} {Approach} with {Earth}},
	volume = {6},
	issn = {2632-3338},
	url = {https://iopscience.iop.org/article/10.3847/PSJ/ae147e},
	doi = {10.3847/PSJ/ae147e},
	language = {en},
	number = {11},
	urldate = {2025-12-03},
	journal = {\psj},
	author = {DeMartini, Joseph V. and Richardson, Derek C. and Murdoch, Naomi and Garcia, Raphaël F. and Schmerr, Nicholas C. and Scheirich, Petr and Ballouz, Ronald-L. and Daca, Adriana},
	month = nov,
	year = {2025},
	pages = {263},
}

@software{devresse2024,
  author = {Devresse, A. and Cornu, N. and Grosheintz-Laval, L. and Awile, O. and de Geus, T. and Pereira, F. and Wolf, M. and HighFive Contributors},
  title = {HighFive - Header-only C++ HDF5 interface},
  year = {2024},
  version = {v2.10.1},
  publisher = {Zenodo},
  doi = {10.5281/zenodo.14272664},
  url = {https://doi.org/10.5281/zenodo.14272664}
}

@article{ferrari2017,
  title={N-body gravitational and contact dynamics for asteroid aggregation},
  author={{Ferrari}, Fabio and {Tasora}, Alessandro and {Masarati}, Pierangelo and {Lavagna}, Michele},
  journal={Multibody System Dynamics},
  volume={39},
  pages={3--20},
  year={2017},
  publisher={Springer}
}

@ARTICLE{ferrari2020,
       author = {{Ferrari}, Fabio and {Lavagna}, Mich{\`e}le and {Blazquez}, Emmanuel},
        title = "{A parallel-GPU code for asteroid aggregation problems with angular particles}",
      journal = {\mnras},
         year = 2020,
        month = feb,
       volume = {492},
       number = {1},
        pages = {749-761},
          doi = {10.1093/mnras/stz3458},
archivePrefix = {arXiv},
       eprint = {1912.04197},
 primaryClass = {astro-ph.EP},
       adsurl = {https://ui.adsabs.harvard.edu/abs/2020MNRAS.492..749F}
}

@ARTICLE{ferrari&tanga2020,
       author = {{Ferrari}, F. and {Tanga}, P.},
        title = "{The role of fragment shapes in the simulations of asteroids as gravitational aggregates}",
      journal = {\icarus},
         year = 2020,
        month = nov,
       volume = {350},
          eid = {113871},
        pages = {113871},
          doi = {10.1016/j.icarus.2020.113871},
archivePrefix = {arXiv},
       eprint = {2005.14032},
 primaryClass = {astro-ph.EP},
       adsurl = {https://ui.adsabs.harvard.edu/abs/2020Icar..35013871F}
}

@ARTICLE{ferrari&tanga2022,
       author = {{Ferrari}, Fabio and {Tanga}, Paolo},
        title = "{Interior of top-shaped asteroids with cohesionless surface}",
      journal = {\icarus},
         year = 2022,
        month = may,
       volume = {378},
          eid = {114914},
        pages = {114914},
          doi = {10.1016/j.icarus.2022.114914},
       adsurl = {https://ui.adsabs.harvard.edu/abs/2022Icar..37814914F}
}

@article{flores2011,
  title={On the continuous contact force models for soft materials in multibody dynamics},
  author={Flores, Paulo and Machado, Margarida and Silva, Miguel T and Martins, Jorge M},
  journal={Multibody system dynamics},
  volume={25},
  number={3},
  pages={357--375},
  year={2011},
  publisher={Springer}
}

@ARTICLE{fodde2026,
       author = {{Fodde}, Iosto and {Civati}, Lucia Francesca and {Cremasco}, Alessia and {Ferrari}, Fabio},
        title = "{Modelling and characterizing spacecraft─surface interactions on small Solar system bodies}",
      journal = {\mnras},
         year = 2026,
        month = apr,
       volume = {547},
       number = {4},
          eid = {stag513},
        pages = {stag513},
          doi = {10.1093/mnras/stag513},
       adsurl = {https://ui.adsabs.harvard.edu/abs/2026MNRAS.547ag513F}
}

@INPROCEEDINGS{herique2025,
       author = {{Herique}, Alain and {Plettemeier}, Dirk and {Haynes}, Mark and {Michel}, Patrick and {Lazzarin}, Monica and {Raymond}, Carol and {Kofman}, Wlodek and {Roger}, Yves},
        title = "{Radar tomography of asteroid deep interior. JuRa / HERA to Didymos and the Radars to Apophis: status of the instruments}",
    booktitle = {EPSC-DPS Joint Meeting 2025},
         year = 2025,
       volume = {2025},
        month = sep,
          eid = {EPSC-DPS2025-463},
        pages = {EPSC-DPS2025-463},
          doi = {10.5194/epsc-dps2025-463},
       adsurl = {https://ui.adsabs.harvard.edu/abs/2025epsc.conf..463H}
}

@ARTICLE{hirabayashi2015_core,
       author = {{Hirabayashi}, Masatoshi and {S{\'a}nchez}, Diego Paul and {Scheeres}, Daniel J.},
        title = "{Internal Structure of Asteroids Having Surface Shedding Due to Rotational Instability}",
      journal = {\apj},
         year = 2015,
        month = jul,
       volume = {808},
       number = {1},
          eid = {63},
        pages = {63},
          doi = {10.1088/0004-637X/808/1/63},
archivePrefix = {arXiv},
       eprint = {1506.03354},
 primaryClass = {astro-ph.EP},
       adsurl = {https://ui.adsabs.harvard.edu/abs/2015ApJ...808...63H}
}

@ARTICLE{hirabayashi2015_cohesion,
       author = {{Hirabayashi}, Masatoshi},
        title = "{Failure modes and conditions of a cohesive, spherical body due to YORP spin-up}",
      journal = {\mnras},
         year = 2015,
        month = dec,
       volume = {454},
       number = {2},
        pages = {2249-2257},
          doi = {10.1093/mnras/stv2017},
archivePrefix = {arXiv},
       eprint = {1508.06913},
 primaryClass = {astro-ph.EP},
       adsurl = {https://ui.adsabs.harvard.edu/abs/2015MNRAS.454.2249H}
}

@article{holsapple2006,
	title = {Tidal disruptions: {A} continuum theory for solid bodies},
	volume = {183},
	issn = {0019-1035},
	shorttitle = {Tidal disruptions},
	url = {https://www.sciencedirect.com/science/article/pii/S0019103506001072},
	doi = {10.1016/j.icarus.2006.03.013},
	number = {2},
	urldate = {2025-11-07},
	journal = {Icarus},
	author = {Holsapple, Keith A. and Michel, Patrick},
	month = aug,
	year = {2006},
	pages = {331--348},
}

@article{holsapple2008,
	title = {Tidal disruptions: {II}. {A} continuum theory for solid bodies with strength, with applications to the {Solar} {System}},
	volume = {193},
	issn = {0019-1035},
	shorttitle = {Tidal disruptions},
	url = {https://www.sciencedirect.com/science/article/pii/S0019103507004538},
	doi = {10.1016/j.icarus.2007.09.011},
	number = {1},
	urldate = {2025-11-07},
	journal = {Icarus},
	author = {Holsapple, Keith A. and Michel, Patrick},
	month = jan,
	year = {2008},
	pages = {283--301},
}

@ARTICLE{granvik2016,
       author = {{Granvik}, Mikael and {Morbidelli}, Alessandro and {Jedicke}, Robert and {Bolin}, Bryce and {Bottke}, William F. and {Beshore}, Edward and {Vokrouhlick{\'y}}, David and {Delb{\`o}}, Marco and {Michel}, Patrick},
        title = "{Super-catastrophic disruption of asteroids at small perihelion distances}",
      journal = {\nat},
         year = 2016,
        month = feb,
       volume = {530},
       number = {7590},
        pages = {303-306},
          doi = {10.1038/nature16934},
       adsurl = {https://ui.adsabs.harvard.edu/abs/2016Natur.530..303G}
}

@article{granvik2024,
	title = {Tidal {Disruption} of {Near}-{Earth} {Asteroids} during {Close} {Encounters} with {Terrestrial} {Planets}},
	volume = {960},
	issn = {2041-8205},
	url = {https://doi.org/10.3847/2041-8213/ad151b},
	doi = {10.3847/2041-8213/ad151b},
	language = {en},
	number = {2},
	urldate = {2025-11-06},
	journal = {The Astrophysical Journal Letters},
	author = {Granvik, Mikael and Walsh, Kevin J.},
	month = jan,
	year = {2024},
	note = {Publisher: The American Astronomical Society},
	pages = {L9},
}

@article{jiang2015,
	title = {A novel three-dimensional contact model for granulates incorporating rolling and twisting resistances},
	volume = {65},
	issn = {0266-352X},
	url = {https://www.sciencedirect.com/science/article/pii/S0266352X14002390},
	doi = {https://doi.org/10.1016/j.compgeo.2014.12.011},
	journal = {Computers and Geotechnics},
	author = {Jiang, Mingjing and Shen, Zhifu and Wang, Jianfeng},
	year = {2015},
	pages = {147--163},
}

@ARTICLE{jura2003,
       author = {{Jura}, M.},
        title = "{A Tidally Disrupted Asteroid around the White Dwarf G29-38}",
      journal = {\apjl},
         year = 2003,
        month = feb,
       volume = {584},
       number = {2},
        pages = {L91-L94},
          doi = {10.1086/374036},
archivePrefix = {arXiv},
       eprint = {astro-ph/0301411},
 primaryClass = {astro-ph},
       adsurl = {https://ui.adsabs.harvard.edu/abs/2003ApJ...584L..91J}
}

@ARTICLE{korycansky2006,
       author = {{Korycansky}, D.~G. and {Asphaug}, Erik},
        title = "{Low-speed impacts between rubble piles modeled as collections of polyhedra}",
      journal = {\icarus},
         year = 2006,
        month = apr,
       volume = {181},
       number = {2},
        pages = {605-617},
          doi = {10.1016/j.icarus.2005.10.028},
       adsurl = {https://ui.adsabs.harvard.edu/abs/2006Icar..181..605K}
}

@ARTICLE{korycansky2009,
       author = {{Korycansky}, D.~G. and {Asphaug}, Erik},
        title = "{Low-speed impacts between rubble piles modeled as collections of polyhedra, 2}",
      journal = {\icarus},
         year = 2009,
        month = nov,
       volume = {204},
       number = {1},
        pages = {316-329},
          doi = {10.1016/j.icarus.2009.06.006},
       adsurl = {https://ui.adsabs.harvard.edu/abs/2009Icar..204..316K}
}

@INPROCEEDINGS{lazzarin2025,
       author = {{Lazzarin}, Monica and {Michel}, Patrick and {Kueppers}, Michael and {Green}, Simon and {Tortora}, Paolo and {Ulamec}, Stephan and {Baptiste Vincent}, Jean and {Abell}, Paul and {Sugita}, Seiji and {Martino}, Paolo},
        title = "{RAMSES: A European rendezvous mission to study tidal effects on the Near-Earth Asteroid Apophis during its 2029 close encounter with the Earth}",
    booktitle = {EPSC-DPS Joint Meeting 2025},
         year = 2025,
       volume = {2025},
        month = sep,
          eid = {EPSC-DPS2025-806},
        pages = {EPSC-DPS2025-806},
          doi = {10.5194/epsc-dps2025-806},
       adsurl = {https://ui.adsabs.harvard.edu/abs/2025epsc.conf..806L}
}

@ARTICLE{li2021,
       author = {{Li}, Daohai and {Mustill}, Alexander J. and {Davies}, Melvyn B.},
        title = "{Accretion of tidally disrupted asteroids on to white dwarfs: direct accretion versus disc processing}",
      journal = {\mnras},
         year = 2021,
        month = dec,
       volume = {508},
       number = {4},
        pages = {5671-5686},
          doi = {10.1093/mnras/stab2949},
archivePrefix = {arXiv},
       eprint = {2106.00441},
 primaryClass = {astro-ph.EP},
       adsurl = {https://ui.adsabs.harvard.edu/abs/2021MNRAS.508.5671L}
}

@article{marohnic2023,
	title = {An {Efficient} {Numerical} {Approach} to {Modeling} the {Effects} of {Particle} {Shape} on {Rubble}-pile {Dynamics}},
	volume = {4},
	issn = {2632-3338},
	url = {https://iopscience.iop.org/article/10.3847/PSJ/ad0467/meta},
	doi = {10.3847/PSJ/ad0467},
	language = {en},
	number = {12},
	urldate = {2025-11-06},
	journal = {\psj},
	author = {{Marohnic}, Julian C. and {DeMartini}, Joseph V. and {Richardson}, Derek C. and {Zhang}, Yun and {Walsh}, Kevin J.},
	month = dec,
	year = {2023},
	note = {Publisher: IOP Publishing},
	pages = {245},
}

@ARTICLE{marohnic2026,
       author = {{Marohnic}, Julian C. and {Richardson}, Derek C. and {Walsh}, Kevin J. and {DeMartini}, Joseph V.},
        title = "{Effect of Irregular Particle Shape in Simulations of Tidal Disruption and Reaccumulation of Small Solar System Bodies}",
      journal = {\psj},
         year = 2026,
        month = mar,
       volume = {7},
       number = {3},
          eid = {72},
        pages = {72},
          doi = {10.3847/PSJ/ae4908},
       adsurl = {https://ui.adsabs.harvard.edu/abs/2026PSJ.....7...72M}
}

@ARTICLE{michel2022,
       author = {{Michel}, Patrick and {K{\"u}ppers}, Michael and {Campo Bagatin}, Adriano and {Carry}, Benoit and {Charnoz}, S{\'e}bastien and {de Leon}, Julia and {Fitzsimmons}, Alan and {Gordo}, Paulo and {Green}, Simon F. and {H{\'e}rique}, Alain and {Juzi}, Martin and {Karatekin}, {\"O}zg{\"u}r and {Kohout}, Tomas and {Lazzarin}, Monica and {Murdoch}, Naomi and {Okada}, Tatsuaki and {Palomba}, Ernesto and {Pravec}, Petr and {Snodgrass}, Colin and {Tortora}, Paolo and {Tsiganis}, Kleomenis and {Ulamec}, Stephan and {Vincent}, Jean-Baptiste and {W{\"u}nnemann}, Kai and {Zhang}, Yun and {Raducan}, Sabina D. and {Dotto}, Elisabetta and {Chabot}, Nancy and {Cheng}, Andy F. and {Rivkin}, Andy and {Barnouin}, Olivier and {Ernst}, Carolyn and {Stickle}, Angela and {Richardson}, Derek C. and {Thomas}, Cristina and {Arakawa}, Masahiko and {Miyamoto}, Hirdy and {Nakamura}, Akiko and {Sugita}, Seiji and {Yoshikawa}, Makoto and {Abell}, Paul and {Asphaug}, Erik and {Ballouz}, Ronald-Louis and {Bottke}, William F. and {Lauretta}, Dante S. and {Walsh}, Kevin J. and {Martino}, Paolo and {Carnelli}, Ian},
        title = "{The ESA Hera Mission: Detailed Characterization of the DART Impact Outcome and of the Binary Asteroid (65803) Didymos}",
      journal = {\psj},
         year = 2022,
        month = jul,
       volume = {3},
       number = {7},
          eid = {160},
        pages = {160},
          doi = {10.3847/PSJ/ac6f52},
       adsurl = {https://ui.adsabs.harvard.edu/abs/2022PSJ.....3..160M}
}

@ARTICLE{michel2025,
       author = {{Michel}, Patrick and {K{\"u}ppers}, Michael and {Fitzsimmons}, Alan and {Green}, Simon and {Lazzarin}, Monica and {Ulamec}, Stephan and {Abell}, Paul and {Sugita}, Seiji and {Campo Bagatin}, Adriano and {Carry}, Benoit and {Charnoz}, S{\'e}bastien and {de Le{\'o}n}, Julia and {Ferrari}, Fabio and {H{\'e}rique}, Alain and {Jutzi}, Martin and {Karatekin}, {\"O}zg{\"u}r and {Kohout}, Tomas and {Murdoch}, Naomi and {Okada}, Tatsuaki and {Palomba}, Ernesto and {Pravec}, Petr and {Raducan}, Sabina and {Snodgrass}, Colin and {Tortora}, Paolo and {Vincent}, Jean-Baptiste and {W{\"u}nnemann}, Kai},
        title = "{The Hera Space Mission in the Context of Small Near-Earth Asteroid Missions in the Past, Present and Future}",
      journal = {\ssr},
         year = 2025,
        month = jul,
       volume = {221},
       number = {5},
          eid = {70},
        pages = {70},
          doi = {10.1007/s11214-025-01195-1},
       adsurl = {https://ui.adsabs.harvard.edu/abs/2025SSRv..221...70M}
}

@article{movshovitz2012,
	title = {Numerical modeling of the disruption of {Comet} {D}/1993 {F2} {Shoemaker}-{Levy} 9 representing the progenitor by a gravitationally bound assemblage of randomly shaped polyhedra},
	volume = {759},
	url = {http://arxiv.org/abs/1207.3386},
	doi = {10.1088/0004-637X/759/2/93},
	number = {2},
	journal = {The Astrophysical Journal},
	author = {Movshovitz, Naor and Asphaug, Erik and Korycansky, Donald},
	month = jul,
	year = {2012},
	note = {Publisher: Institute of Physics Publishing},
	pages = {93--93},
}

@ARTICLE{pajola2024,
       author = {{Pajola}, M. and {Tusberti}, F. and {Lucchetti}, A. and {Barnouin}, O. and {Cambioni}, S. and {Ernst}, C.~M. and {Dotto}, E. and {Daly}, R.~T. and {Poggiali}, G. and {Hirabayashi}, M. and {Nakano}, R. and {Epifani}, E. Mazzotta and {Chabot}, N.~L. and {Della Corte}, V. and {Rivkin}, A. and {Agrusa}, H. and {Zhang}, Y. and {Penasa}, L. and {Ballouz}, R. -L. and {Ivanovski}, S. and {Murdoch}, N. and {Rossi}, A. and {Robin}, C. and {Ieva}, S. and {Vincent}, J.~B. and {Ferrari}, F. and {Raducan}, S.~D. and {Campo-Bagatin}, A. and {Parro}, L. and {Benavidez}, P. and {Tancredi}, G. and {Karatekin}, {\"O}. and {Trigo-Rodriguez}, J.~M. and {Sunshine}, J. and {Farnham}, T. and {Asphaug}, E. and {Deshapriya}, J.~D.~P. and {Hasselmann}, P.~H.~A. and {Beccarelli}, J. and {Schwartz}, S.~R. and {Abell}, P. and {Michel}, P. and {Cheng}, A. and {Brucato}, J.~R. and {Zinzi}, A. and {Amoroso}, M. and {Pirrotta}, S. and {Impresario}, G. and {Bertini}, I. and {Capannolo}, A. and {et al.}},
        title = "{Evidence for multi-fragmentation and mass shedding of boulders on rubble-pile binary asteroid system (65803) Didymos}",
      journal = {Nature Communications},
         year = 2024,
        month = jul,
       volume = {15},
          eid = {6205},
        pages = {6205},
          doi = {10.1038/s41467-024-50148-9},
       adsurl = {https://ui.adsabs.harvard.edu/abs/2024NatCo..15.6205P}
}

@PHDTHESIS{stadel2001,
       author = {{Stadel}, Joachim Gerhard},
        title = "{Cosmological N-body simulations and their analysis}",
       school = {University of Washington, Seattle},
         year = 2001,
        month = jan,
       adsurl = {https://ui.adsabs.harvard.edu/abs/2001PhDT........21S}
}

@ARTICLE{raducan&jutzi2022,
       author = {{Raducan}, Sabina D. and {Jutzi}, Martin},
        title = "{Global-scale Reshaping and Resurfacing of Asteroids by Small-scale Impacts, with Applications to the DART and Hera Missions}",
      journal = {\psj},
         year = 2022,
        month = jun,
       volume = {3},
       number = {6},
          eid = {128},
        pages = {128},
          doi = {10.3847/PSJ/ac67a7},
       adsurl = {https://ui.adsabs.harvard.edu/abs/2022PSJ.....3..128R}
}

@ARTICLE{raducan2024a,
       author = {{Raducan}, S.~D. and {Jutzi}, M. and {Cheng}, A.~F. and {Zhang}, Y. and {Barnouin}, O. and {Collins}, G.~S. and {Daly}, R.~T. and {Davison}, T.~M. and {Ernst}, C.~M. and {Farnham}, T.~L. and {Ferrari}, F. and {Hirabayashi}, M. and {Kumamoto}, K.~M. and {Michel}, P. and {Murdoch}, N. and {Nakano}, R. and {Pajola}, M. and {Rossi}, A. and {Agrusa}, H.~F. and {Barbee}, B.~W. and {Syal}, M. Bruck and {Chabot}, N.~L. and {Dotto}, E. and {Fahnestock}, E.~G. and {Hasselmann}, P.~H. and {Herreros}, I. and {Ivanovski}, S. and {Li}, J.-Y. and {Lucchetti}, A. and {Luther}, R. and {Orm{\"o}}, J. and {Owen}, M. and {Pravec}, P. and {Rivkin}, A.~S. and {Robin}, C.~Q. and {S{\'a}nchez}, P. and {Tusberti}, F. and {W{\"u}nnemann}, K. and {Zinzi}, A. and {Epifani}, E. Mazzotta and {Manzoni}, C. and {May}, B.~H.},
        title = "{Physical properties of asteroid Dimorphos as derived from the DART impact}",
      journal = {Nature Astronomy},
         year = 2024,
        month = apr,
       volume = {8},
        pages = {445-455},
          doi = {10.1038/s41550-024-02200-3},
archivePrefix = {arXiv},
       eprint = {2403.00667},
 primaryClass = {astro-ph.EP},
       adsurl = {https://ui.adsabs.harvard.edu/abs/2024NatAs...8..445R}
}

@ARTICLE{raducan2024b,
       author = {{Raducan}, S.~D. and {Jutzi}, M. and {Merrill}, C.~C. and {Michel}, P. and {Zhang}, Y. and {Hirabayashi}, M. and {Mainzer}, A.},
        title = "{Lessons Learned from NASA's DART Impact about Disrupting Rubble-pile Asteroids}",
      journal = {\psj},
         year = 2024,
        month = mar,
       volume = {5},
       number = {3},
          eid = {79},
        pages = {79},
          doi = {10.3847/PSJ/ad29f6},
archivePrefix = {arXiv},
       eprint = {2403.00683},
 primaryClass = {astro-ph.EP},
       adsurl = {https://ui.adsabs.harvard.edu/abs/2024PSJ.....5...79R}
}

@article{richardson1998,
	title = {Tidal {Distortion} and {Disruption} of {Earth}-{Crossing} {Asteroids}},
	volume = {134},
	doi = {10.1006/icar.1998.5954},
	number = {1},
	journal = {Icarus},
	author = {Richardson, Derek C. and Bottke, William F. and Love, Stanley G.},
	month = jul,
	year = {1998},
	note = {Publisher: Academic Press},
	pages = {47--76},
}

@ARTICLE{richardson2000,
       author = {{Richardson}, Derek C. and {Quinn}, Thomas and {Stadel}, Joachim and {Lake}, George},
        title = "{Direct Large-Scale N-Body Simulations of Planetesimal Dynamics}",
      journal = {\icarus},
         year = 2000,
        month = jan,
       volume = {143},
       number = {1},
        pages = {45-59},
          doi = {10.1006/icar.1999.6243},
       adsurl = {https://ui.adsabs.harvard.edu/abs/2000Icar..143...45R}
}

@INCOLLECTION{richardson2002,
       author = {{Richardson}, D.~C. and {Leinhardt}, Z.~M. and {Melosh}, H.~J. and {Bottke}, Jr., W.~F. and {Asphaug}, E.},
        title = "{Gravitational Aggregates: Evidence and Evolution}",
    booktitle = {Asteroids III},
         year = 2002,
       editor = {{Bottke}, Jr., W.~F. and {Cellino}, A. and {Paolicchi}, P. and {Binzel}, R.~P.},
        pages = {501-515},
       adsurl = {https://ui.adsabs.harvard.edu/abs/2002aste.book..501R}
}

@article{roche1847,
  title={Memoirs divers sur l'equilibre d'une mass fluide},
  author={Roche, E},
  journal={Acad. Sci. Lett. Montpelier. Mem. Section Sci.},
  volume={1},
  pages={243},
  year={1847}
}

@ARTICLE{sanchez2012,
  title={DEM simulation of rotation-induced reshaping and disruption of rubble-pile asteroids},
  author={{S{\'a}nchez}, P. and {Scheeres}, D.~J.},
  journal={Icarus},
  volume={218},
  number={2},
  pages={876--894},
  doi={10.1016/j.icarus.2012.01.014},
  year={2012},
  publisher={Elsevier}
}

@ARTICLE{sanchez2014,
       author = {{S{\'a}nchez}, P. and {Scheeres}, D.~J.},
        title = "{The strength of regolith and rubble pile asteroids}",
      journal = {Meteroritics \& Planetary Science},
         year = 2014,
        month = may,
       volume = {49},
       number = {5},
        pages = {788-811},
          doi = {10.1111/maps.12293},
archivePrefix = {arXiv},
       eprint = {1306.1622},
 primaryClass = {astro-ph.EP},
       adsurl = {https://ui.adsabs.harvard.edu/abs/2014M&PS...49..788S}
}

@article{sanchez2016,
title = {Disruption patterns of rotating self-gravitating aggregates: A survey on angle of friction and tensile strength},
journal = {Icarus},
volume = {271},
pages = {453-471},
year = {2016},
issn = {0019-1035},
doi = {https://doi.org/10.1016/j.icarus.2016.01.016},
url = {https://www.sciencedirect.com/science/article/pii/S0019103516000208},
author = {{S{\'a}nchez}, P. and {Scheeres}, D.~J.}
}

@ARTICLE{sanchez2018,
       author = {{S{\'a}nchez}, Paul and {Scheeres}, Daniel J.},
        title = "{Rotational evolution of self-gravitating aggregates with cores of variable strength}",
      journal = {\planss},
         year = 2018,
        month = aug,
       volume = {157},
        pages = {39-47},
          doi = {10.1016/j.pss.2018.04.001},
       adsurl = {https://ui.adsabs.harvard.edu/abs/2018P&SS..157...39S}
}

@ARTICLE{scipy,
  author  = {Virtanen, Pauli and Gommers, Ralf and Oliphant, Travis E. and
            Haberland, Matt and Reddy, Tyler and Cournapeau, David and
            Burovski, Evgeni and Peterson, Pearu and Weckesser, Warren and
            Bright, Jonathan and {van der Walt}, St{\'e}fan J. and
            Brett, Matthew and Wilson, Joshua and Millman, K. Jarrod and
            Mayorov, Nikolay and Nelson, Andrew R. J. and Jones, Eric and
            Kern, Robert and Larson, Eric and Carey, C J and
            Polat, {\.I}lhan and Feng, Yu and Moore, Eric W. and
            {VanderPlas}, Jake and Laxalde, Denis and Perktold, Josef and
            Cimrman, Robert and Henriksen, Ian and Quintero, E. A. and
            Harris, Charles R. and Archibald, Anne M. and
            Ribeiro, Ant{\^o}nio H. and Pedregosa, Fabian and
            {van Mulbregt}, Paul and {SciPy 1.0 Contributors}},
  title   = {{{SciPy} 1.0: Fundamental Algorithms for Scientific
            Computing in Python}},
  journal = {Nature Methods},
  year    = {2020},
  volume  = {17},
  pages   = {261--272},
  adsurl  = {https://rdcu.be/b08Wh},
  doi     = {10.1038/s41592-019-0686-2},
}

@ARTICLE{schunova2014,
       author = {{Schunov{\'a}}, Eva and {Jedicke}, Robert and {Walsh}, Kevin J. and {Granvik}, Mikael and {Wainscoat}, Richard J. and {Haghighipour}, Nader},
        title = "{Properties and evolution of NEO families created by tidal disruption at Earth}",
      journal = {\icarus},
         year = 2014,
        month = aug,
       volume = {238},
        pages = {156-169},
          doi = {10.1016/j.icarus.2014.05.006},
archivePrefix = {arXiv},
       eprint = {1405.4090},
 primaryClass = {astro-ph.EP},
       adsurl = {https://ui.adsabs.harvard.edu/abs/2014Icar..238..156S}
}

@ARTICLE{sridhar1992,
       author = {{Sridhar}, S. and {Tremaine}, S.},
        title = "{Tidal disruption of viscous bodies}",
      journal = {\icarus},
         year = 1992,
        month = jan,
       volume = {95},
       number = {1},
        pages = {86-99},
          doi = {10.1016/0019-1035(92)90193-B},
       adsurl = {https://ui.adsabs.harvard.edu/abs/1992Icar...95...86S}
}

@ARTICLE{sekanina1994,
       author = {{Sekanina}, Z. and {Chodas}, P.~W. and {Yeomans}, D.~K.},
        title = "{Tidal disruption and the appearance of periodic comet Shoemaker-Levy 9.}",
      journal = {\aap},
         year = 1994,
        month = sep,
       volume = {289},
        pages = {607-636},
       adsurl = {https://ui.adsabs.harvard.edu/abs/1994A&A...289..607S}
}

@ARTICLE{sunday2020,
       author = {{Sunday}, Cecily and {Murdoch}, Naomi and {Tardivel}, Simon and {Schwartz}, Stephen R. and {Michel}, Patrick},
        title = "{Validating N-body code CHRONO for granular DEM simulations in reduced-gravity environments}",
      journal = {\mnras},
         year = 2020,
        month = oct,
       volume = {498},
       number = {1},
        pages = {1062-1079},
          doi = {10.1093/mnras/staa2454},
archivePrefix = {arXiv},
       eprint = {2009.10448},
 primaryClass = {astro-ph.EP},
       adsurl = {https://ui.adsabs.harvard.edu/abs/2020MNRAS.498.1062S}
}

@InCollection{Chrono2016,
  Title                    = {Chrono: An open source multi-physics dynamics engine},
  Author                   = {Tasora, A. and Serban, R. and Mazhar, H. and Pazouki, A. and Melanz, D. and Fleischmann, J. and Taylor, M. and Sugiyama, H. and Negrut, D.},
  Booktitle                = {High Performance Computing in Science and Engineering – Lecture Notes in Computer Science},
  Publisher                = {Springer},
  Year                     = {2016},
  Editor                   = {Kozubek, T.},
  Pages                    = {19--49}
}

@misc{ChronoGitHub,
  author = {{Project Chrono Development Team}},
  title = {Chrono: {An Open Source Framework for the Physics-Based Simulation of Dynamic Systems}},
  url = {https://github.com/projectchrono/chrono},
  month = may,
  year = {2017},
  note = {Accessed: 2026-08-12},
}

@article{vaghi2025,
	title = {Contact dynamics investigation towards microgravity experiment for asteroid-related scenarios},
	issn = {1573-272X},
	url = {https://doi.org/10.1007/s11044-025-10091-z},
	doi = {10.1007/s11044-025-10091-z},
	language = {en},
	urldate = {2026-01-05},
	journal = {Multibody System Dynamics},
	author = {Vaghi, Samuele and Fodde, Iosto and Panicucci, Paolo and Cremasco, Alessia and Ferrari, Fabio},
	month = nov,
	year = {2025},
}

@article{walsh2006,
	title = {Binary near-{Earth} asteroid formation: {Rubble} pile model of tidal disruptions},
	volume = {180},
	issn = {0019-1035},
	shorttitle = {Binary near-{Earth} asteroid formation},
	url = {https://www.sciencedirect.com/science/article/pii/S0019103505003210},
	doi = {10.1016/j.icarus.2005.08.015},
	number = {1},
	urldate = {2025-11-06},
	journal = {Icarus},
	author = {Walsh, Kevin J. and Richardson, Derek C.},
	month = jan,
	year = {2006},
	pages = {201--216},
}

@ARTICLE{walsh2018,
       author = {{Walsh}, Kevin J.},
        title = "{Rubble Pile Asteroids}",
      journal = {\araa},
         year = 2018,
        month = sep,
       volume = {56},
        pages = {593-624},
          doi = {10.1146/annurev-astro-081817-052013},
archivePrefix = {arXiv},
       eprint = {1810.01815},
 primaryClass = {astro-ph.EP},
       adsurl = {https://ui.adsabs.harvard.edu/abs/2018ARA&A..56..593W}
}

@ARTICLE{wimarsson2024,
       author = {{Wimarsson}, John and {Xiang}, Zhen and {Ferrari}, Fabio and {Jutzi}, Martin and {Madeira}, Gustavo and {Raducan}, Sabina D. and {S{\'a}nchez}, Paul},
        title = "{Rapid formation of binary asteroid systems post rotational failure: A recipe for making atypically shaped satellites}",
      journal = {\icarus},
         year = 2024,
        month = oct,
       volume = {421},
          eid = {116223},
        pages = {116223},
          doi = {10.1016/j.icarus.2024.116223},
archivePrefix = {arXiv},
       eprint = {2407.15543},
 primaryClass = {astro-ph.EP},
       adsurl = {https://ui.adsabs.harvard.edu/abs/2024Icar..42116223W}
}

@ARTICLE{wimarsson2025,
       author = {{Wimarsson}, John and {Ferrari}, Fabio and {Jutzi}, Martin},
        title = "{The diverse shapes of binary asteroid satellites born from sub-escape-velocity moonlet mergers}",
      journal = {\aap},
         year = 2025,
        month = dec,
       volume = {704},
          eid = {A29},
        pages = {A29},
          doi = {10.1051/0004-6361/202555914},
       adsurl = {https://ui.adsabs.harvard.edu/abs/2025A&A...704A..29W}
}

@ARTICLE{yu2014,
       author = {{Yu}, Yang and {Richardson}, Derek C. and {Michel}, Patrick and {Schwartz}, Stephen R. and {Ballouz}, Ronald-Louis},
        title = "{Numerical predictions of surface effects during the 2029 close approach of Asteroid 99942 Apophis}",
      journal = {\icarus},
         year = 2014,
        month = nov,
       volume = {242},
        pages = {82-96},
          doi = {10.1016/j.icarus.2014.07.027},
archivePrefix = {arXiv},
       eprint = {1408.0168},
 primaryClass = {astro-ph.EP},
       adsurl = {https://ui.adsabs.harvard.edu/abs/2014Icar..242...82Y}
}

@ARTICLE{zhang2017,
       author = {{Zhang}, Yun and {Richardson}, Derek C. and {Barnouin}, Olivier S. and {Maurel}, Clara and {Michel}, Patrick and {Schwartz}, Stephen R. and {Ballouz}, Ronald-Louis and {Benner}, Lance A.~M. and {Naidu}, Shantanu P. and {Li}, Junfeng},
        title = "{Creep stability of the proposed AIDA mission target 65803 Didymos: I. Discrete cohesionless granular physics model}",
      journal = {\icarus},
         year = 2017,
        month = sep,
       volume = {294},
        pages = {98-123},
          doi = {10.1016/j.icarus.2017.04.027},
archivePrefix = {arXiv},
       eprint = {1703.01595},
 primaryClass = {astro-ph.EP},
       adsurl = {https://ui.adsabs.harvard.edu/abs/2017Icar..294...98Z}
}

@ARTICLE{zhang2018,
       author = {{Zhang}, Yun and {Richardson}, Derek C. and {Barnouin}, Olivier S. and {Michel}, Patrick and {Schwartz}, Stephen R. and {Ballouz}, Ronald-Louis},
        title = "{Rotational Failure of Rubble-pile Bodies: Influences of Shear and Cohesive Strengths}",
      journal = {\apj},
         year = 2018,
        month = apr,
       volume = {857},
       number = {1},
          eid = {15},
        pages = {15},
          doi = {10.3847/1538-4357/aab5b2},
       adsurl = {https://ui.adsabs.harvard.edu/abs/2018ApJ...857...15Z}
}

@ARTICLE{zhang2020a,
       author = {{Zhang}, Yun and {Lin}, Douglas N.~C.},
        title = "{Tidal fragmentation as the origin of 1I/2017 U1 (`Oumuamua)}",
      journal = {Nature Astronomy},
         year = 2020,
        month = apr,
       volume = {4},
        pages = {852-860},
          doi = {10.1038/s41550-020-1065-8},
archivePrefix = {arXiv},
       eprint = {2004.07218},
 primaryClass = {astro-ph.EP},
       adsurl = {https://ui.adsabs.harvard.edu/abs/2020NatAs...4..852Z}
}

@article{zhang2020b,
	title = {Tidal distortion and disruption of rubble-pile bodies revisited - {Soft}-sphere discrete element analyses},
	volume = {640},
	copyright = {© Y. Zhang and P. Michel 2020},
	issn = {0004-6361, 1432-0746},
	url = {https://www.aanda.org/articles/aa/abs/2020/08/aa37856-20/aa37856-20.html},
	doi = {10.1051/0004-6361/202037856},
	language = {en},
	urldate = {2025-11-06},
	journal = {\aap},
	author = {Zhang, Yun and Michel, Patrick},
	month = aug,
	year = {2020},
	note = {Publisher: EDP Sciences},
	pages = {A102},
}

@ARTICLE{zhang2021,
       author = {{Zhang}, Yun and {Michel}, Patrick and {Richardson}, Derek C. and {Barnouin}, Olivier S. and {Agrusa}, Harrison F. and {Tsiganis}, Kleomenis and {Manzoni}, Claudia and {May}, Brian H.},
        title = "{Creep stability of the DART/Hera mission target 65803 Didymos: II. The role of cohesion}",
      journal = {\icarus},
         year = 2021,
        month = jul,
       volume = {362},
          eid = {114433},
        pages = {114433},
          doi = {10.1016/j.icarus.2021.114433},
       adsurl = {https://ui.adsabs.harvard.edu/abs/2021Icar..36214433Z}
}

\begin{appendix}
\nolinenumbers

\section{Upgrades to GRAINS}\label{appendix:grains_changes}

Here, we outline recent upgrades to our numerical code GRAINS \citep{ferrari2017,ferrari2020}, which have significantly improved its performance, enabling planetary-scale simulations. With GPU acceleration, we can now run simulations with tens of thousands of particles, extending our realistic in-simulation timescales from days to weeks.

\subsection{New version of Project CHRONO}

The open-source physics engine Project CHRONO \citep{Chrono2016} has seen major enhancements since it was originally implemented in GRAINS \citep{ferrari2017,ferrari2020}, using its version 3.0. Recently, the GRAINS development team has upgraded GRAINS to CHRONO version 9.0.1. Here, we list the most relevant changes that this brought. The full changelogs of CHRONO can be found on their GitHub page \citep{ChronoGitHub}.

\begin{itemize}
    \item The CHRONO::MULTICORE module, which parallelises contact detection and characterisation. Compared to the now deprecated CHRONO::PARALLEL used in the old version of GRAINS, it also has faster parallelisation when parsing particle data structures.  
    \item More detailed collision shape models for imported meshes and generated convex hulls, along with increased contact force precision from FLOAT32 to FLOAT64, which leads to more accurate contact force calculations. 
    \item The non-linear Flores contact model, which has an added hysteresis damping term that leads to a more continuous repulsive force during collisions compared to the Hertzian model \citep{flores2011,sunday2020}.
    \item User-defined contact force calculation, where the user can supplement the default contact calculations with a custom class for additional forces such as cohesion.
    \item Ability to implement a custom contact force reporter (see Appendix~\ref{appendix:pressure}).
\end{itemize}

\subsection{Binary file format}

We developed a new module for writing and reading data from GRAINS simulations, based on the C++ library HighFive, which implements the HDF5 format \citep{devresse2024}. The properties of all particles are separated into meta and frame data. The former represents constant information such as masses, moments of inertia, densities, and vertex positions relative to the particle barycentre. The frame data type contains step-specific information such as position and velocity. All data are stored in long arrays of length $N_\mathrm{p}$ as INT32, FLOAT32 or FLOAT64, depending on the precision needed for the specific data. For example, all velocities and positions of barycentres and convex hull vertices are stored in FLOAT64, while the quaternions that represent the rotational state of a given particle are stored as FLOAT32. Furthermore, we also rewrote the structure of our simulation output to allow for mixing of particle types by using a shape flag, as previous versions of GRAINS used shape-specific output and input formats. Currently, the shape flags are as follows:

\begin{enumerate}
    \setcounter{enumi}{0}
    \item no shape: does not have a convex hull or sphere information, rotational state fixed;
    \item sphere: defined in terms of a radius and density, tracks rotation;
    \item convex hull: defined by hull vertices relative to the barycentre and density, tracks rotation. 
\end{enumerate}

This means we can now mix spheres and convex hulls in a given setup. Furthermore, going from the old text-based module to one that uses binary files, we are able to retain more detailed information regarding the properties of each particle while also speeding up writing and reading operations during simulations and analysis. For a simulation of typical particle number and duration, we see a reduction in total storage size by a factor of ten.

\subsection{Particle generation}

 We further made improvements to the way we generate particles to properly resolve the individual vertices of irregular particle meshes relative to the particle centre-of-mass, even at distances of several planetary radii. In the old version of GRAINS, we would always initiate the vertices of a given particle at its target position. To elaborate, a particle generated at position $\vec{r}_p$ was assigned a convex hull vertex with the absolute position $\vec{r}_{v,\mathrm{abs}} = \vec{r}_p + \vec{r}_{v,\mathrm{loc}}$, where $\vec{r}_{v,\mathrm{loc}}$ is its position in the local particle frame of reference. For distances $\vec{r}_p \gg \vec{r}_{v,\mathrm{loc}}$, CHRONO would then have difficulties resolving the difference in position for different vertices belonging to the same convex hull when calculating its shape and collision models. 
 
 After identifying this issue, we now always initiate a convex hull at the origin, such that the initial position of the vertex is $\vec{r}_{v,\mathrm{abs}} = \vec{r}_{v,\mathrm{loc}}$, and then proceed to move it to the target position after we have calculated the shape and collision models. As vertices are always provided in the local frame of reference in CHRONO for resolving contacts and rotational operations, there are no longer any resolution issues, even for cases where $\vec{r}_p \gg \vec{r}_{v,\mathrm{loc}}$.

\subsection{Updating the gravitational force module}

The original version of our CUDA GPU-accelerated gravitational force calculator was presented in \citet{ferrari2020}, and it is largely based on the implementation by \citet{burtscher2011}. The structure remains the same, with five kernels that carry out the following tasks:

\begin{enumerate}
    \item calculate the bounding box of the system;
    \item octree construction;
    \item compute the centre-of-mass for each node;
    \item sort bodies by distance;
    \item compute gravitational forces.
\end{enumerate}

The upgrade mainly concerns the transition to newer releases of CUDA than version 9. Many of the built-in tools for performing memory and task synchronisation across threads and warps that we used in older versions of GRAINS are deprecated in more recent versions of CUDA, which are required for newer NVIDIA hardware. This restricted many of the recent GRAINS studies to regular $\mathcal{O}(N^2)$ force calculation, substantially reducing performance \citep[e.g.][]{wimarsson2024}. The module is now fully integrated with CUDA versions 10 and upwards, employing modern synchronisation functions across GPU threads with safer parallelisation. Moreover, the old Barnes-Hut module was built to use single-precision floats, which provided enough information for asteroid-scale systems. However, for the planetary-scale simulations explored in this study, the distances between a particle, $i$, and the planet, $|\vec{r}_{p,i}|$, can be more than a factor $10^7$ greater than the distance between two adjacent particles, $i$ and $j$, given by $|\vec{r}_{p,i} - \vec{r}_{p,j}|$. In turn, the system becomes highly sensitive to small changes in the gravitational force caused by the planetary perturber, and a wrongly calculated acceleration can generate an exaggerated overlap between two particles, which can lead to an overestimated contact force and diverging numerical solutions.

\section{Weibull fitting}\label{appendix:weibull}

\begin{table}[!ht]
    \centering
    \caption{Weibull distribution parameters for the fits in Fig.~\ref{fig:final_masses_agg1}.}
    \begin{tabular}{ccccccc}
    \hline\hline
    $v_\infty$ & $q$ & $N_\mathrm{frags}$ & $k$ & $\lambda$ & $D_{KS,\mathrm{obs}}$ & $p_{KS}$ \\
    \hline
    0 & 1.1 & 49 & 0.9198 & 0.0195 & 0.120 & 0.078 \\
    0 & 1.2 & 43 & 0.8898 & 0.0216 & 0.113 & 0.178 \\
    0 & 1.3 & 39 & 1.0176 & 0.0256 & 0.091 & 0.548 \\
    0 & 1.4 & 38 & 0.8865 & 0.0246 & 0.105 & 0.360 \\
    0 & 1.5 & 36 & 0.8548 & 0.0256 & 0.148 & 0.040 \\
    0 & 1.6 & 32 & 0.7677 & 0.0264 & 0.156 & 0.044 \\
    0 & 1.7 & 23 & 0.8702 & 0.0407 & 0.160 & 0.144 \\
    0 & 1.8 & 15 & 0.8314 & 0.0612 & 0.231 & 0.026 \\
    0 & 1.9 & 14 & 0.4651 & 0.0305 & 0.240 & 0.028 \\
    0 & 2.0 & 7 & 0.5282 & 0.0840 & 0.204 & 0.550 \\
    2 & 1.1 & 43 & 0.9623 & 0.0227 & 0.109 & 0.216 \\
    2 & 1.2 & 37 & 0.8568 & 0.0248 & 0.146 & 0.054 \\
    2 & 1.3 & 30 & 0.9196 & 0.0319 & 0.138 & 0.146 \\
    2 & 1.4 & 30 & 0.9156 & 0.0319 & 0.110 & 0.486 \\
    2 & 1.5 & 25 & 1.0437 & 0.0404 & 0.185 & 0.016 \\
    2 & 1.6 & 21 & 0.9194 & 0.0458 & 0.205 & 0.026 \\
    2 & 1.7 & 19 & 0.6755 & 0.0406 & 0.172 & 0.152 \\
    2 & 1.8 & 16 & 0.5537 & 0.0375 & 0.177 & 0.198 \\
    2 & 1.9 & 9 & 0.5832 & 0.0718 & 0.185 & 0.552 \\
    4 & 1.1 & 34 & 0.8111 & 0.0263 & 0.172 & 0.012 \\
    4 & 1.2 & 26 & 0.8887 & 0.0362 & 0.111 & 0.574 \\
    4 & 1.3 & 25 & 0.8600 & 0.0370 & 0.157 & 0.120 \\
    4 & 1.4 & 21 & 0.9648 & 0.0468 & 0.112 & 0.656 \\
    4 & 1.5 & 13 & 1.1303 & 0.0793 & 0.251 & 0.020 \\
    4 & 1.6 & 13 & 0.8846 & 0.0729 & 0.178 & 0.338 \\
    4 & 1.7 & 9 & 0.6163 & 0.0789 & 0.186 & 0.532 \\
    4 & 1.8 & 5 & 0.4858 & 0.1069 & 0.248 & 0.510 \\
    6 & 1.1 & 25 & 0.7712 & 0.0342 & 0.153 & 0.160 \\
    6 & 1.2 & 18 & 1.0197 & 0.0558 & 0.185 & 0.086 \\
    6 & 1.3 & 13 & 0.7897 & 0.0698 & 0.286 & 0.004 \\
    6 & 1.4 & 10 & 1.4035 & 0.1074 & 0.180 & 0.486 \\
    6 & 1.5 & 5 & 2.5409 & 0.2258 & 0.336 & 0.082 \\
    8 & 1.1 & 13 & 1.2260 & 0.0813 & 0.119 & 0.904 \\
    8 & 1.2 & 10 & 0.9885 & 0.0994 & 0.327 & 0.006 \\
    8 & 1.3 & 6 & 1.2731 & 0.1775 & 0.275 & 0.196 \\
    \hline
    \end{tabular}
    \label{tab:weibull_fits}
\end{table}

\noindent Analysing the resulting distributions of masses in our simulations, we work under the null hypothesis that they are Weibull in nature. Given that our values, $m$, are always larger than zero, we get the following probability density function:

\begin{equation}\label{eq:weibull}
    f(m;\lambda,k) = \frac{k}{\lambda}\left( \frac{m}{\lambda}\right)^{k-1}e^{-(m/\lambda)^k},
\end{equation}

\noindent where $k>0$ is the shape parameter and $\lambda$ is the scale parameter. 

For each test, we bootstrapped $N_\mathrm{boot}=500$ synthetic samples, drawing $N_\mathrm{frag}$ values from the initial, `observed' cumulative Weibull distribution and performed a new fit to the data. The Kolmogorov-Smirnov (KS) test statistic for each bootstrapped sample, $D_{KS,\mathrm{boot}}$, was then compared to the observed KS test statistic, $D_{KS,\mathrm{obs}}$. The resulting p-value is then calculated as

\begin{equation}\label{eq:p_value_KS}
    p_{KS} = \frac{1}{N_\mathrm{boot}} \sum\limits_{b=1}^{N_\mathrm{boot}} \mathds{1} \{ D_{KS,\text{boot}} \geq D_{KS,\text{obs}} \},
\end{equation}

\noindent where $\mathds{1}$ is the indicator function. 

\section{Stress and dissipation}\label{appendix:pressure}

The stress analysis performed in Sect.~\ref{section:stress_profiles} is enabled by our implementation of a customised CHRONO callback reporter (`contact reporter'). The contact reporter gives a more detailed, contact-level characterisation of the system, but incurs additional computational overhead. Pressure is computed each time step via a per-particle Cauchy stress tensor, $\sigma$ \citep{sanchez2012, zhang2017}. For each contact involving particle \(i\), the contact force is accumulated as

\begin{equation}
\boldsymbol{\sigma}_i
=
\frac{1}{V_i}
\sum_{j \in \mathcal{C}_i}
\mathbf{f}_{ij}\otimes \boldsymbol{\ell}_{ij},
\end{equation}

\noindent
where \(V_i\) is the particle volume, \(\mathbf{f}_{ij}\) is the contact force, and \(\boldsymbol{\ell}_{ij}\) is the branch vector from the particle centre to the contact point. The per-particle pressure $p_i$ as reported in Fig.~\ref{fig:pressure_profiles} is here taken as the negative one-third trace of its stress tensor,

\begin{equation}
p_i = -\frac{1}{3}\operatorname{tr}\left(\boldsymbol{\sigma}_i\right) = -\frac{1}{3}\left(\sigma_{i,xx}+\sigma_{i,yy}+\sigma_{i,zz}\right).
\end{equation}

We track the cumulative dissipation shown in Fig.~\ref{fig:cumulative_dissipation} from the net contact-force power acting on the particles. At time step $n$, the instantaneous dissipative power is estimated from the CHRONO-reported net contact force on each particle, $\mathbf{F}_{c,i}^{(n)}$, and the particle velocity, $\mathbf{v}_i^{(n)}$, as

\begin{equation}
P_{\mathrm{diss}}^{(n)}
=
\sum_i
\max\left[
0,\,
-\mathbf{F}_{c,i}^{(n)}\cdot \mathbf{v}_i^{(n)}
\right],
\end{equation}
\noindent
where only positive values are retained so that the diagnostic measures contact work removed from particle motion. The per-step dissipated energy is

\begin{equation}
\Delta E_{\mathrm{diss}}^{(n)}
=
P_{\mathrm{diss}}^{(n)}\Delta t,
\end{equation}
\noindent
and the cumulative dissipation through time step $N$ is

\begin{equation}
E_{\mathrm{diss}}^{\mathrm{cum}}(N)
=
\sum_{n=1}^{N}
\Delta E_{\mathrm{diss}}^{(n)}
=
\sum_{n=1}^{N}
\Delta t
\sum_i
\max\left[
0,\,
-\mathbf{F}_{c,i}^{(n)}\cdot \mathbf{v}_i^{(n)}
\right].
\end{equation}

This quantity should therefore be interpreted as a diagnostic measure of contact-force work loss, rather than a complete decomposition of the underlying DEM dissipation mechanisms. A higher-resolution accounting of the individual numerical dissipation channels and their mapping onto physical loss mechanisms is left to future work.

\end{appendix}

\end{document}